\documentclass[apj,iop]{emulateapj}

\usepackage{amsmath}
\usepackage{tabularx}
\usepackage{natbib}
\usepackage{float}
\usepackage[colorlinks=true,linkcolor=blue,citecolor=blue]{hyperref}%
\usepackage[all]{hypcap}
\usepackage{lineno}
\usepackage{footnote}
\usepackage{academicons}

\newcommand{\teff}{$T_{\rm eff}$}

\newcommand{\co}{CO}
\newcommand{\meth}{CH$_4$}
\newcommand{\amon}{NH$_3$}
\newcommand{\cotwo}{CO$_2$} 
\newcommand{\water}{H$_2$O}
\newcommand{\phos}{PH$_3$}
\newcommand{\pq}{$P_Q$}

\newcommand{\tchem}{$t_{\rm chem}$}
\newcommand{\tmix}{$t_{\rm mix}$}
\newcommand{\kzz}{$K_{zz}$}

\newcommand{\orcid}[1]{\href{https://orcid.org/#1}{\includegraphics[width=10pt]{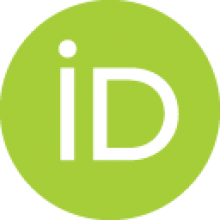}}}

\shorttitle{Atmospheric Mixing and Chemistry in Brown Dwarfs}
\shortauthors{Mukherjee et al.}
\graphicspath{{./}{figures/}}

\begin{document}

\title{Probing the Extent of Vertical Mixing in Brown Dwarf Atmospheres with Disequilibrium Chemistry}

\email{samukher@ucsc.edu}

\author{Sagnick Mukherjee$^{1}$ \orcid{0000-0003-1622-1302}, Jonathan J. Fortney$^{1}$ \orcid{0000-0002-9843-4354}, Natasha E. Batalha$^{2}$ \orcid{0000-0003-1240-6844}, Theodora Karalidi$^{3}$ \orcid{0000-0001-7356-6652}, Mark S. Marley$^{4}$ \orcid{0000-0002-5251-2943}, Channon Visscher$^{5}$ \orcid{0000-0001-6627-6067}, Brittany E. Miles$^{1}$ \orcid{0000-0002-5500-4602}, and Andrew J. I. Skemer$^{1}$ \orcid{0000-0001-6098-3924}}
\affiliation{{$^1$}Department of Astronomy and Astrophysics, University of California, Santa Cruz, CA 95064, USA \\ 
{$^2$} NASA Ames Research Center, MS 245-3, Moffett Field, CA 94035, USA \\
{$^3$}Department of Physics, University of Central Florida, 4111 Libra Dr, Orlando, FL 32816, USA\\
{$^4$} Lunar and Planetary Laboratory, The University of Arizona, Tucson, AZ 85721, USA\\
{$^5$} Dordt University, Sioux Center IA; Space Science Institute, Boulder, CO, USA\\}

\begin{abstract}

Evidence of disequilibrium chemistry due to vertical mixing in the atmospheres of many T and Y-dwarfs has been inferred due to enhanced mixing ratios of {\co} and reduced {\amon}. Atmospheric models of planets and brown dwarfs typically parameterize this vertical mixing phenomenon with the vertical eddy diffusion coefficient, {\kzz}. While {\kzz} can perhaps be approximated in the convective regions in the atmosphere with mixing length theory, in radiative regions the strength of vertical mixing is uncertain by many orders of magnitude. With a new grid of self-consistent 1D model atmospheres from \teff\ of 400 - 1000 K, computed with a new radiative-convective equilibrium python code \texttt{PICASO 3.0}, we aim to assess how molecular abundances and corresponding spectra can be used as a probe of depth-dependent {\kzz}.  At a given surface gravity, we find non-monotonic behavior in the CO abundance as a function of \teff, as chemical abundances are sometimes quenched in either of two potential atmospheric convective zones, or quenched in either of two possible radiative zones.  The temperature structure and chemical quenching behavior also changes with gravity.  We compare our models with available near-infrared and M-band spectroscopy of several T and Y-dwarfs and assess their atmospheric vertical mixing profiles.  We also compare to color-magnitude diagrams and make predictions for JWST spectra.  This work yields new constraints, and points the way to significant future gains, in determining {\kzz}, a fundamental atmospheric parameter in substellar atmospheres, with significant implications for chemistry and cloud modeling.

\end{abstract}

\section{Introduction}\label{sec:intro}
\subsection{Molecular abundances and non-equilibrium chemistry}
The astronomical community is now ready to observe the atmospheres of a diverse range of exoplanets and brown dwarfs in unprecedented detail with the {\it James Webb Space Telescope} (JWST) \citep{Gardner06jwst,JWSTERO}. These atmospheres are complex systems with a range of physical processes like radiative/convective energy transport, chemical reaction networks, clouds/haze formation, and dynamical processes like vertical mixing. Upcoming high signal--to--noise (SNR) spectra from JWST for these substellar atmospheres will provide us with a unique opportunity to probe these physical and chemical processes and compare to state--of--the--art theoretical models. Therefore, it is crucial to develop and update these models by including necessary complex processes and assessing how uncertainties in physical and chemical parameters affect the structure, abundances, and spectra of these atmospheres.

One such complex process in substellar atmospheres is atmospheric dynamics.  Dynamics is  a three-dimensional process. In the case of solar system planets, atmospheric dynamics can be directly studied with spatially and time resolved imaging or spectroscopy. Although, we cannot spatially resolve brown dwarfs and exoplanets, dynamical processes can still be studied by their impact on molecular abundances and condensate clouds, which affect their spectra and lightcurves.  Surface inhomogeneities can be probed via rotationally modulate lightcurves \citep{woow14,cushing16,Karalidi15,vos22}, often in two dimensions when observed at a single wavelength.  Spectra, which probe a range of atmospheric pressures, provide a potential view of dynamics in the additional dimension of atmosphere depth.  Processes like convection, the breaking of gravity waves, and convective overshoot can impact both the chemical and cloud structure of these atmospheres by causing net vertical transport of atmospheric gases and particles across many atmospheric scale heights \citep{freytag10,parmentier13,bordwell18,tan22}.

Dynamical processes can also alter the abundances of atmospheric gases in the visible atmosphere by transporting gas molecules through several scale heights, on a timescale faster than chemical reactions are able to locally equilibrate to their new surroundings. This leads to departures from the chemical equilibrium, often assumed as a starting point in exoplanet and brown dwarf atmosphere models \citep[e.g.,][]{bobcat}. The dynamics induced transport of gases and particles along the radial dimension of these atmospheres is called ``vertical mixing". Vertical mixing can lead to disequilibrium chemical abundances in substellar atmospheres for gases like {\co}, {\meth}, {\water}, {\amon}, {\cotwo}, PH$_3$, and HCN \citep{fortney20,tsai17,tsai21,moses11}. Because most of these gases are major sources of opacities in substellar atmospheres ({\water}, {\meth}, {\co}, etc.),  their spectra can be altered significantly. 

Vertical mixing also directly affects cloud formation through it's effect on molecules like H$_2$O in Y-dwarfs and colder exoplanets \citep{morley14,ackerman2001cloud,gao2020aerosol}. Vertical mixing transports condensable vapor from the deeper atmosphere to lower pressures and temperatures, where they condense. Moreover, mixing counteracts gravitational settling, and therefore helps to dictate the size of cloud particles that can remain aloft. As a result, the atmospheric vertical mixing significantly impacts the cloud opacity by controlling the cloud vertical extent and particle size distribution. As both the chemical and cloud structure of these atmospheres impact radiative and convective energy transport, vertical mixing also affects the temperature structure of substellar atmospheres.

There is already decades-deep literature on observations and modeling of disequilibrium chemistry in Jupiter and Saturn \citep{beer75,larson78,prinn77,larson77,fegley85,visscher2010icarus}, and in T-dwarfs and early Y-dwarfs \citep{Fegley96,noll97,oppenheimer98,saumon03,golimowski04,Hubeny07,geballe09,sorahana2012,leggett12,visscher2010apj,visscher2011,Zahnle14,Miles20}. We briefly summarize the key developments in the field in \S\ref{sec:modeling_approach}. But, more recently, \citet{Miles20} obtained M-band spectra of four late T and early Y-dwarfs and used archival M-band spectra of two T-dwarfs and a Y-dwarf to infer the strength of vertical mixing in their atmospheres as a function of {\teff}. They determined the {\co} abundance in each of these objects by fitting models to spectra. 1D brown dwarf atmospheric models with equilibrium chemistry were post-processed to include the effect of disequilibrium chemistry due to vertical mixing to interpret the observed {\co} abundances and infer the vigor of vertical mixing in these atmospheres. This demonstrated that measurements of chemical abundances of various gases along with 1D forward models can be used to constrain vertical mixing in substellar atmospheres. However, disequilibrium chemistry can lead to significant changes in the atmospheric state of these objects compared to equilibrium chemistry calculations due to the feedback of the modified chemical abundances of gases on the atmospheric temperature structure \citep{Hubeny07,karilidi21,Philips20}.

The \citet{Miles20} paper and new modeling work points to the possibility of using upcoming high SNR JWST spectra of brown dwarfs to probe their atmospheric structure and atmospheric chemistry in a much more detailed way.  The rationale of our study here is to specifically examine how differences in the strength of vertical mixing, and how it is typically parametrized in convective and radiative atmospheric regions, gives rise to distinct differences in model spectra.  Comparison with measured spectra now and in the future can then be used to assess the mixing speed in convective and radiative regions, yielding novel constraints on atmospheric physics.

\subsection{Modeling Approaches}\label{sec:modeling_approach}
Mixing in substellar atmospheres is a 3-dimensional process. Both 2D and 3D models have been used to study vertical mixing in hotter brown dwarfs and hot Jupiters \citep{tan22,tan21,parmentier13,freytag10,bordwell18}. However, a full exploration of the relevant parameter space of brown dwarf mixing properties, is currently not feasible with 3D models. Moreover, these 3D models often include simplified or gray radiative transfer, so in detail the calculated temperature structures yield model spectra that are not yet an adequate match to measured spectra \citep{tan21,mukherjee21}. Chemical equilibrium is also often assumed in GCMs \citep[e.g.,][]{lee2021,showman09} but some models also treat chemical species within a passive transport framework \citep[e.g.,][]{parmentier13,Komacek_2019,tan22} to explore the effects of atmospheric mixing on chemical structure of substellar atmospheres.

1D self-consistent models, which iterate to a radiative-convective equilibrium (RCE) solution, given specified  atmospheric physics and chemistry, include many of the complex physical processes of real atmospheres and can be used to produce high SNR model observables.  Given the relatively fast computational times, they are important in exploring wide ranges of parameter space \citep[e.g.][]{Marley96,burrows97,marley1999thermal,Burrows_2003,fortney08,morley14,gandhi17,malik17,marley21}. 
Although such models only include one dimension, the effects of dynamics can still be included.  This is typically done by approximating vertical mixing as a diffusive process, and parameterizing  this diffusive ``speed' with a single vertical eddy diffusion coefficient -- {\kzz} \citep{allen81} which has dimensions of [L]$^2$/[T], where [L] and [T] represent length and time, respectively. However, theoretical and observational constraints of {\kzz}, and its dependence on temperature and pressure, is uncertain by several orders of magnitude today \citep{fortney20,Miles20,Philips20}. 


The vigor of vertical mixing could well have a complex behavior, even within the atmosphere of a single object.  Brown dwarf atmospheres have convective and radiative zones. In the convective zones mixing is expected to be comparatively efficient with higher values of {\kzz}, compared to the radiative zones which are stable to convection.  Theoretically, {\kzz} has been approximated with mixing length theory in the convective zones of these atmospheres \citep[e.g.,][]{ackerman01,smith1998,gierasch85}. However, this approximation is still uncertain due to uncertainties in key parameters in the mixing length theory like the mixing length itself.  Theoretical understanding of {\kzz} is even more primitive in the case of radiative zones.  Typically, {\kzz} is varied by a factor of a \emph{million} or more, with its effects on the temperature structure and atmospheric abundances inadequately explored. This poor knowledge of {\kzz} leaves large gaps in our understanding of vertical mixing on chemistry, clouds, and temperature structure.

Previously, several authors  computed RCE models that treated non-equilibrium chemical abundances self-consistently with the atmospheric temperature structure \citep[e.g.][]{Hubeny07,Philips20,karilidi21}.  However, these models were all computed either within the framework of a constant {\kzz} value in the atmosphere, with an exploration of  {\kzz} varied across many orders of magnitude \citep[e.g.][]{karilidi21,Philips20}, or with {\kzz} computed using mixing length theory in the convective zones and constant {\kzz} in the radiative zones \citep[e.g.][]{Hubeny07}.  We build on these works, but aim for a more precise assessment of how a changing {\kzz} within a given atmosphere, due to convective- and radiative-zone mixing, alters atmospheric abundances and spectra.  The output from 3D GCMs show that {\kzz} can vary by several orders of magnitude with pressure in a substellar atmosphere \citep{freytag10,parmentier13,tan22}, but this has not yet been implemented within a self-consistent RCE model.

Observations from \citet{Miles20}  showed that T-dwarfs with {\teff} between 400--800 K have much smaller {\kzz} than expected from convective mixing. They also found a factor of 100 \emph{increase} in the inferred {\kzz} in these brown dwarfs  when {\teff} \emph{decreased} below $\sim$ 400 K. \citet{Miles20} suggested an explanation for this behavior if in the warm objects with {\teff} between 400-800 K, gases are ``quenched'' within a radiative part of the atmosphere (lower {\kzz}), but in a convective part of the atmosphere in objects with {\teff}$<400 $K. Therefore, already available observations demand models with higher complexity than a constant {\kzz} approach. This is especially needed to prepare these models adequately to interpret the much higher quality data expected from JWST.

In this work, we enhance the models presented in \citet{karilidi21} by inclusion of temperature- and pressure- dependant {\kzz} profiles within the 1D RCE framework. We use this model to explore the complex interconnections between {\kzz}, the temperature structure, molecular abundances, and spectra in late T and early Y-dwarfs. We use this model and already available ground-based spectroscopic data of six brown dwarfs to  address the following questions,
 \begin{enumerate}
      
      \item How does a self-consistent treatment of vertical mixing affect the $T(P)$ profile?
      
      \item Does a self-consistent treatment of {\kzz} cause changes in the radiative and convective zone locations and depths in brown dwarfs compared to equilibrium chemistry models?
      
      \item In what objects can we expect quenching of gases like {\co}, {\meth} and {\amon} to occur in radiative zones instead of convective zones?
      
      
      \item Is photometry and spectroscopy of brown dwarfs between 1-14 $\mu$m sensitive to the {\kzz} in the radiative and convective zone?
      
      \item How much can we learn about the atmospheric properties of late T-dwarfs by applying  self-consistent disequilibrium chemistry models on already available infrared spectral data?
      
      \item Can JWST be used to accurately infer the vertical mixing strength in radiative zones and convective zones of brown dwarfs?
      
 \end{enumerate}
 
We briefly discuss our new Python based atmospheric model in \S\ref{sec:model}. We present our results in \S\ref{sec:results} followed by discussion of our results in \S\ref{sec:disc} and conclude in \S\ref{sec:sum}

\section{Modeling Sub-stellar Atmospheres}\label{sec:model}
We have adapted the Fortran based \texttt{EGP} sub-stellar atmospheric model to Python. This new Python version of the \texttt{EGP} code is open-source and available as a part of the widely used Python exoplanet atmospheric modeling tool \texttt{PICASO} \citep{batalha19}, as a part of the \texttt{PICASO 3.0} \citep{Mukherjee22} release. The numerical and functional details of the code along with a series of benchmarks with previously available models are presented in \citet{Mukherjee22}. The \texttt{EGP} model has been used by our group for substellar atmospheres for over two decades \citep{Marley96,Marley_2012,fortney2005comp,fortney2007planetary,fortney08,morley14,marley21,karilidi21}. Recently, \citet{karilidi21} updated the Fortran based \texttt{EGP} code with the capability to include constant {\kzz} and its impact on chemistry self-consistently. We have converted this version of the code in Python and have modified it by including the capability to treat pressure-dependent variable {\kzz} in substellar atmospheres. Here we briefly describe the model methodology used in this work and we refer the reader to \citet{Mukherjee22} for a more detailed discussion of the numerical methodology of our models.

\subsection{Python Based implementation of \texttt{EGP} with Disequilibrium Chemistry}\label{sec:deq_py}

We divide the 1D model atmosphere in 90 pressure layers (91 layer boundaries or levels) with logarithmically-spaced pressure values. From an initial guess temperature-pressure ($T(P)$) profile, the model iterates on the $T(P)$ profile, atmospheric chemistry, and radiative/convective energy transport using a Newton-Rhapson method, yielding temperature corrections at each pressure level. These iterations continue until it finds the converged atmospheric state in which radiative-convective equilibrium is satisfied throughout the atmosphere. Self-consistent modeling of effects like disequilibrium chemistry due to vertical mixing requires simultaneous calculation of gas abundances and opacities with the iteration on the $T(P)$ profile instead of interpolation on a precomputed grid.

Each chemical reaction in the atmosphere proceeds with a characteristic timescale {\tchem} which is a strong function of temperature, pressure, and the chemical composition of the atmosphere. \citet{Zahnle14} used 1D chemical kinetics models to parameterize {\tchem} expression for the major O-, C-, H-, and N-bearing reactions in substellar atmospheres. These reactions often are a part of a large network of chemical reactions with many chemical species but still can be represented by simpler net reactions between a few species of interest. The net chemical reactions between O-, C-, H- and N-bearing gases we consider here are,

\begin{gather*}
{\rm CH_4 + H_2O \rightleftarrows CO + 3H_2} \\
{\rm CO + H_2O \rightleftarrows CO_2 + H_2} \\
{\rm 2NH_3 \rightleftarrows N_2 + 3H_2} \\
{\rm CH_4 + NH_3 \rightleftarrows HCN + 3H_2} \\
{\rm 2CO + N_2 + 3H_2 \rightleftarrows 2HCN + 2H_2O}\\
{\rm CO + NH_3 \rightleftarrows HCN + H_2O}\\
\end{gather*}

Vertical mixing in the atmosphere also proceeds with a typical timescale, which is associated with the eddy diffusion coefficient {\kzz},
\begin{equation}\label{eq:tmix}
    t_{\rm mix} = \dfrac{H^2}{K_{\rm zz}}
\end{equation}
where $H$ is the local scale height of that atmospheric layer. \tchem\ is generally much shorter than \tmix\ in high pressure, high temperature regions of the atmosphere where gas abundances are expected to follow chemical equilibrium. But as the atmosphere gets colder at lower pressures  \tchem\ rises exponentially and becomes much longer than {\tmix}. The pressure at which the intersection between the two timescales occurs is called the ``quench pressure" or ``quench level."  At all pressure levels lower than the quench pressure (higher up the atmosphere) atmospheric abundances for the relevant chemical compounds take on the abundances at the relevant quench pressure, rather than those of equilibrium chemistry at the local $P$ and $T$.  In the brown dwarf context this behavior was first predicted by \citet{Fegley96} and observed by \citet{noll97}. 

Within our model, the $T(P)$ profile is used to calculate the {\kzz} as a function of each model layer's pressure. {\kzz} in the convective zone is calculated with mixing length theory \citep{gierasch85},

\begin{equation}\label{eq:kzz_con}
 K_{zz}= \dfrac{H}{3}{\left(\dfrac{L}{H}\right)}^{{4}/{3}}{\left(\dfrac{RF}{\mu{\rho_a}c_p}\right)}^{{1}/{3}}
\end{equation}

where $H$ is the local scale height of the atmosphere, $L$ is the turbulent mixing length, $R$ is the universal gas constant, $\mu$ is the mean molecular weight of the atmosphere, $\rho_a$ is the atmospheric density, $c_p$ is the atmospheric specific heat at constant pressure and $F$ is the convective heat flux. The convective heat flux is given by the difference between $\sigma{T_{\rm eff}}^4$ and the radiative flux at each layer within the convective zones of the atmosphere. The maximum convective heat flux would then be $\sigma{T_{\rm eff}}^4$. Here the convective heat flux is calculated self-consistently in each layer. The turbulent mixing length $L$ is set to be equal to the scale height ($H$) in our models. But in order to explore scenarios where convective mixing is less effective, we  also simulate cases where the mixing length is smaller than the atmospheric scale height $H$ by a factor of 10. Note that this turbulent mixing length ($L$) was only changed in Equation \ref{eq:kzz_con} and not in Equation \ref{eq:tmix}. The mixing length in Equation \ref{eq:tmix} remained the same as the local atmospheric scale height throughout this study. For further background, we refer the reader to previous work \citep[e.g.][]{visscher2011,smith1998,bordwell18,BEZARD2002,visscher2010icarus} exploring the relationship between mixing length and the scale height in Equation \ref{eq:tmix}.



Over the past several years various parameterizations of {\kzz} in the radiative zones of substellar atmospheres have been discussed in the literature \citep[e.g.,][]{wang15,zhang18,parmentier13}.  As a starting point, we  use the parametrization from \citet{moses21},

\begin{equation}\label{eq:kzz_rad}
    K_{zz} = \dfrac{5\times10^{8}}{\sqrt{P_{bar}}}\left(\dfrac{H_{\rm 1 mbar}}{{620}\,  {\rm km}}\right)\left(\dfrac{T_{\rm eff}}{1450\, {\rm K}}\right)^{4}
\end{equation}
where $P_{\rm bar}$ is the pressure in bars and $H_{\rm 1 mbar}$ is the atmospheric scale height at 10$^{-3}$ bars. The 5$\times$10$^8$/$\sqrt{P}$ dependence was estimated by \citet{parmentier13} using tracer transport in 3D GCM for hot Jupiter HD209458b. This dependence was found to be valid for this object between 10$^{-6}$-1 bars. The P$^{-0.5}$ dependence captures the increase of {\kzz} with lowering pressure in the radiative zone as mixing can be efficient in low pressure (P $\sim$ 10$^{-4}$ bars) parts of the radiative atmosphere due to mechanisms like gravity wave breaking \citep{freytag10}. The factor $H_{\rm 1 mbar}T_{\rm eff}^4$ was also included in the parametrization as this factor shows good correlation with the measured {\kzz} (via photochemical models) of solar system planets (including Jupiter, Saturn, Uranus, Neptune, Earth, and Venus) and exoplanets \citep{moses05,zhang18,moses21}. But, as the $T(P)$ structure of brown dwarfs can be quite different from irradiated planets, we also explore deviations from this parametrization where this relation is increased or decreased by factors of 100 to explore the relevant parameter space. As the $T(P)$ profile iterates towards the converged solution, the {\kzz} in convective zones, given by Equation \ref{eq:kzz_con} and the {\kzz} in radiative zones, given by Equation \ref{eq:kzz_rad}, evolve as well. 
The mixing timescale in each atmospheric layer is then calculated in every iteration from this {\kzz} using Equation \ref{eq:tmix}. The quench pressures of each of these chemical species are found by comparing \tmix\ with \tchem\ expressions from \citet{Zahnle14}.  The abundances of {\co}, {\meth}, {\amon}, {\cotwo}, {\water} and HCN are allowed to follow equilibrium chemistry at pressures higher than their relevant ``quench pressure'' and their abundances are quenched above the quench pressure level. But as the $T(P)$ profile iterates to the converged solution, the quench pressure of various gases also change at each step as both the {\tchem} and {\tmix} are dependent on the $T(P)$ profile of the atmosphere. This requires the RCE model to include ``on--the--fly" calculations of both chemistry and opacities in order to include disequilibrium chemistry self-consistently in the calculations.

As in \cite{karilidi21} we use the ``on--the--fly mixing" method of opacities following the methodology of \citet{amundsen17}. 
In this work we focus on the quenching of {\co}, {\meth}, {\water}, {\amon}, N$_2$, HCN and PH$_3$.  Of these, only {\co}, {\meth}, {\water}, and {\amon} are major opacity sources.  Therefore, we mix the correlated-k opacities of \co, \meth, \water, and {\amon} with the correlated-k opacities of all the other sets of gases which follow equilibrium chemistry as the volume mixing ratio of these gases evolve due to quenching with the $T(P)$ profile of the atmosphere. As in \cite{karilidi21}, given inaccuracies inherent in the flexible \citet{amundsen17} resort-rebin approach necessitates the use of a denser grid of 661 wavelength bins, compared to the 196 used for our equilibrium chemistry models \citep{marley21}.

We  benchmark the code in two ways: 1) by setting {\kzz}= 0 and finding that the on--the--fly mixing method still converges to the equilibrium chemistry solutions from \citet{marley21}, and 2) by testing the resulting models against results obtained by \citet{Philips20} with non-zero but constant {\kzz} values. The benchmarking results are presented in \citet{Mukherjee22}. The code results matches well with both models.   In future work the code will be further enhanced via coupling with 1D chemical kinetics model \texttt{VULCAN} \citep{tsai17,tsai21} so that it can self-consistently treat both vertical mixing and photochemistry. This Python based code will be open-sourced as a part of the \texttt{PICASO 3.0} package. Next, we discuss results obtained by applying this code to model disequilibrium chemistry and characterize mixing strengths in late T and early Y-dwarf atmospheres.

\section{Results}\label{sec:results}
We performed a detailed four dimensional parameter space exploration of how varied vigor of vertical mixing in brown dwarf atmospheres affects the atmospheric state and spectra. We produced a grid of models by varying four physical parameters -- {\teff}, gravity, radiative zone {\kzz}, and convective zone mixing length $L$. The gravity has been varied across five values 316, 562, 1000, 1780 and 3160 ms$^{-2}$ ($\log(g)$ from 4.5 to 5.5 with an interval of 0.25 dex). The {\teff} range is 400 K to 1000 K with an interval of 25 K.

To explore the role of radiative zone {\kzz}, in our grids we multiply Equation \ref{eq:kzz_rad} by a multiplicative factor of $\times$ 100, $\times$ 1, and $\times$ 0.01. We refer to these parameter values as ``Moses$\times$100", ``Moses", and ``Moses/100", respectively, after \citet{moses21}. The mixing length parameter in the convective zones ($L$ in Equation \ref{eq:kzz_con}) has been varied between the scale height $H$ and $0.1\times{H}$ to explore scenarios where mixing in the convective zone is less efficient than the typically assumed mixing length theory in 1D models. Furthermore, for reasons discussed in \S\ref{sec:disc}, we ran two sets of models -- one where PH$_3$ is quenched given its appropriate {\tchem} and another where PH$_3$ follows equilibrium chemistry while other molecules are quenched. This parameter space has a total of 5$\times$25$\times$3$\times$2$\times$2$=$1500 self-consistent radiative-convective equilibrium models. All of these models are cloud free and assume solar metallicity and C/O ratio, following \citet{marley21}. We haven't included clouds in our models as at the temperature range of our grid (400 -- 1000 K), clouds are not expected to be a significant opacity source in the upper atmosphere. We present our key findings from this grid of models below beginning with the feedback of disequilibrium chemistry on $T(P)$ profiles of brown dwarfs.

\subsection{Effect of Disequilibrium Chemistry on the $T(P)$ profile of brown dwarfs}\label{sec:tp_effect}

\begin{figure*}
  \centering
  \includegraphics[width=1\textwidth]{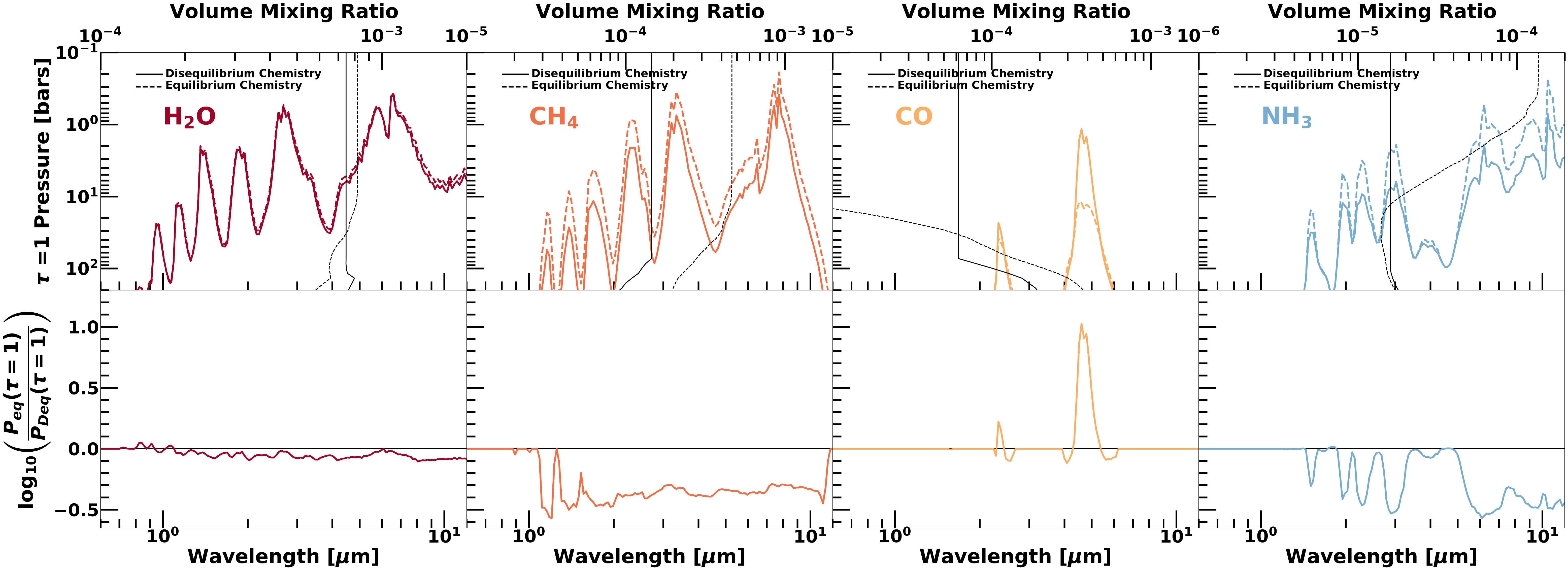}
  \caption{{\bf Top panels} show the $\tau= 1$ pressure levels for specific gaseous absorbers in a model T-dwarf with \teff = 700 K and log(g) = 5.0 in the presence and absence of quenching due to vertical mixing. $\tau= 1$ pressure levels for {\water}, {\meth}, {\co}, and {\amon} are shown in the first, second, third and fourth columns, respectively. The equilibrium chemistry pressure levels are shown in dashed lines whereas the disequilibrium chemistry pressure levels are shown in solid lines.  The volume mixing ratio as a function of pressure of each gas in the presence (solid lines) and absence (dashed lines) of vertical mixing induced quenching are also shown with black lines in the top x-axis of each of the top four panels. {\bf Bottom panels} shows the logarithm of the pressure ratio of these $\tau= 1$  levels with equilibrium chemistry and disequilibrium chemistry for the gaseous absorbers. The zero mark represents no change in these pressure levels. Positive values represents an increase in absorption in the disequilibrium chemistry models compared to equilibrium chemistry models. {\bf Main Point} -- Disequilibrium chemistry leads to large opacity changes in the atmosphere.} 
\label{fig:fig1}
\end{figure*}

Disequilibrium chemistry leads to enhanced {\co}, depleted {\meth}, {\amon}, and {\water} abundances in T and Y-dwarfs. This leads to major changes in the wavelength-dependent optical depths in the atmosphere. The top four panels of Figure \ref{fig:fig1} shows the $\tau= 1$ pressure levels, for specific gases, for a T-dwarf with {\teff} = 700 K and log(g) = 5.0 in the presence and absence of quenching due to vertical mixing. These $\tau = 1$ pressure levels are strongly wavelength dependent. Here {\kzz} has been assumed to follow Equation \ref{eq:kzz_con} in the convective zones and Equation \ref{eq:kzz_rad} in radiative zones. The disequilibrium chemistry pressure levels are shown in solid lines and the corresponding equilibrium chemistry  levels are shown as dotted lines. The pressure-dependant volume mixing ratios of these gases under chemical equilibrium and disequilibrium are also shown in the top x-axis in each of the top four panels with dashed and solid black lines, respectively.

The bottom panels in Figure \ref{fig:fig1} depicts the log of the ratio in the $\tau = 1$ pressure level between these two models. Values greater than zero represent shallower depth probed, due to an increase in an absorber of interest, for instance {\co} in yellow in the third column, in particular at 4.5-5 ${\mu}m$. There are also significant changes due to depleted {\amon} and {\meth}. There is a more modest effect on {\water}. Overall, since major opacity sources {\water}, {\amon}, {\meth} are depleted, and they absorb across a broad wavelength range, one sees deeper, to a higher pressure, in a disequilibrium chemistry model.

\begin{figure*}
  \centering
  \includegraphics[width=0.9\textwidth]{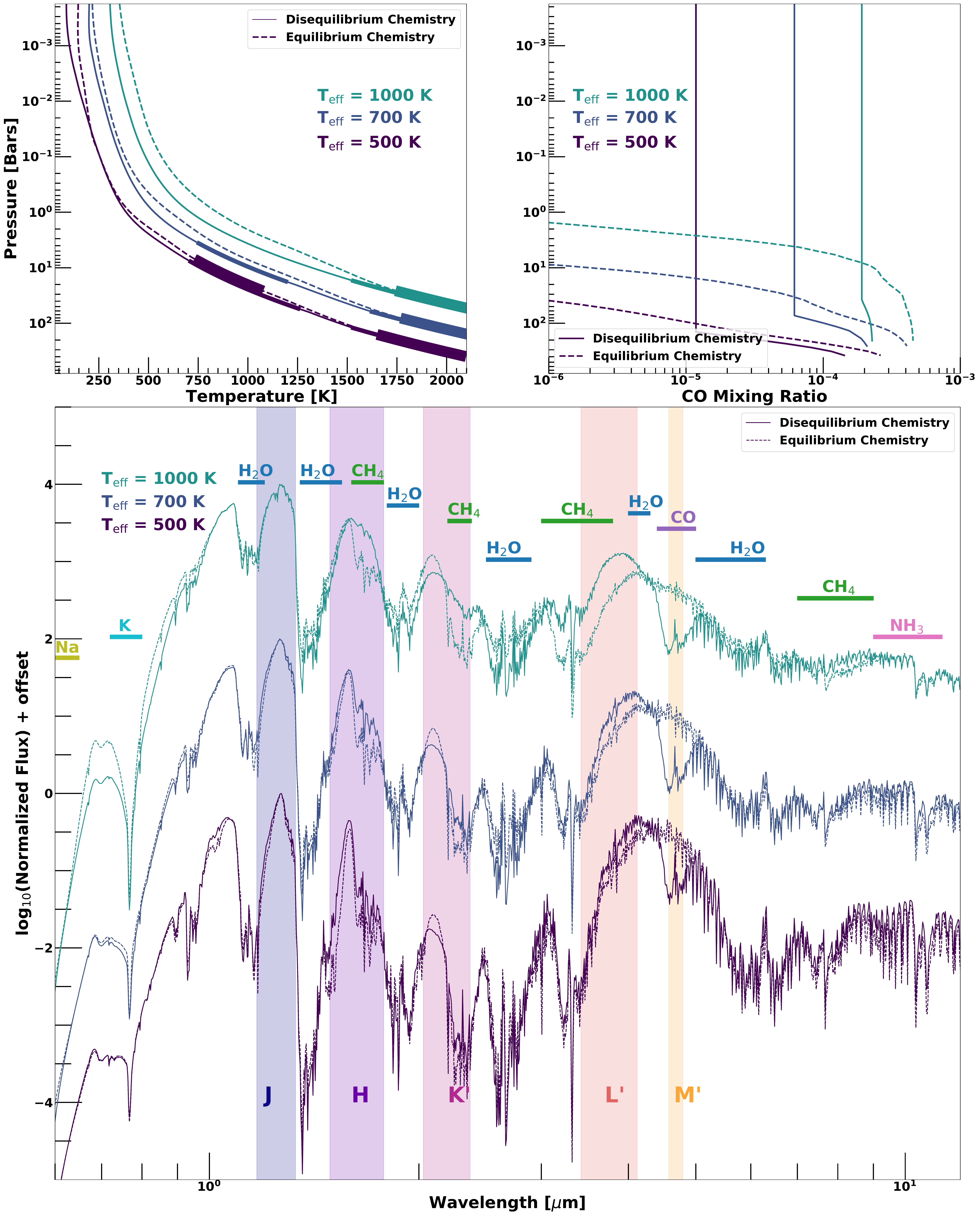}
  \caption{{\bf Top left panel} $T(P)$ profile of brown dwarf models with log(g) = 5 and {\teff} of 500, 700, and 1000 K. The solid $T(P)$ profiles assume disequilibrium chemistry self-consistently whereas the dotted profiles show equilibrium chemistry calculations. The convective zones are shown with thicker lines for each $T(P)$ profile. To differentiate the convective zones of the equilibrium chemistry models from the disequilibrium chemistry models, they are shown with much thicker lines compared to the convective zones of disequilibrium chemistry models. {\bf Top right panel} Plotted are the {\co} volume mixing ratio profiles for these brown dwarf models. The solid lines show the mixing ratio with disequilibrium chemistry whereas the dotted lines show calculations from equilibrium chemistry. {\bf Bottom panel} shows the normalized thermal emission spectra of these brown dwarfs. The solid lines show the disequilibrium chemistry results while dotted spectra represent the equilibrium chemistry calculations. The J, H, K', L', and M' bands are also shown as shaded regions of different colors. {\bf Main Point}-- Self- consistent treatment of disequilibrium chemistry leads to large changes in $T(P)$ profile, abundances, and thermal emission spectra.}
\label{fig:fig2}
\end{figure*}

A manifestation of these effects are shown in Figure \ref{fig:fig2}, which shows temperature structure (upper left), CO mixing ratios (upper right), and spectra (bottom) three models at {\teff} = 500, 700, and 1000 K and log(g) = 5. Solid lines include   induced quenching and dashed lines represent equilibrium chemistry.  The {\kzz} for these models  evolved self-consistently following mixing length theory in the convective zones and  followed Equation \ref{eq:kzz_rad} in the radiative zones.

Disequilibrium abundances mainly lead to colder $T(P)$ profiles compared to equilibrium chemistry models as can be seen in Figure \ref{fig:fig2}. In the overall optically thinner disequilibrium chemistry atmosphere (see Figure \ref{fig:fig1}) one must see down to a higher pressure to obtain a given $T$. This effect was also seen in \citet{karilidi21} for constant {\kzz} models.

Volume mixing ratio profiles of {\co} for each of these models are shown in the top right panel of Figure \ref{fig:fig2}.  As described earlier, disequilibrium chemistry greatly enhances the {\co} abundances above quench level.  However, even deep down, below the quench level, there are important {\co} abundance differences due to differences in the converged $T(P)$ profile. As equilibrium chemistry models have hotter $T(P)$ profiles, at a given pressures, than disequilibrium chemistry models, they also have higher {\co} abundances. This drives home the point that a self-consistent treatment of disequilibrium chemistry  (as in \S\ref{sec:deq_py}) is needed rather than post-processing self-consistent equilibrium chemistry models with quenched abundances. Even if one obtains the correct quench pressures during post-processing equilibrium chemistry models, one would calculate a physically inconsistent and higher abundance of {\co} at the upper atmospheres than self-consistent disequilibrium chemistry models.

The thermal emission spectra of these models are shown in Figure \ref{fig:fig2} bottom panel. The flux from these objects were normalized by their maximum flux and are plotted from the hottest model (\teff =1000 K) at the top to the coldest model (\teff =500 K) at the bottom.  Large differences in outgoing thermal flux can be seen in the H, K, L and M bands between the equilibrium chemistry and disequilibrium chemistry model spectra. Comparing these differences with the differences in the $\tau = 1$ pressure levels in Figure \ref{fig:fig1} makes it clear that the differences in the H, K, and L bands arises due to changes in the {\meth} abundance profiles.  Another major difference is in M-band. \co\ enhancement and modest {\water} depletion leads to a major suppression of thermal flux between 4.5-5 $\mu$m in the disequilibrium models.

Although \amon\ abundances show large changes between equilibrium chemistry and disequilibrium chemistry models, it still is not the major opacity source at the shorter IR wavelengths. It becomes a major opacity source at 11 $\mu$m and causes a distinct feature in the spectra. Figure \ref{fig:fig2} bottom panels show a significant change in this {\amon} feature  due to disequilibrium chemistry. Quenching of {\amon} causes this feature to become slightly less prominent compared to chemical equilibrium models.

Other carbon and nitrogen bearing molecules like HCN, N$_2$ and CO$_2$ are also quenched due to {\kzz}. But these molecules do not have readily apparent absorption features in the infrared at these spectral resolutions in a solar metallicity mixture. N$_2$ is essentially transparent in solar metallicity mixtures and detecting {\cotwo} is very difficult from the ground but JWST, like AKARI \citep{sorahana2012}, can detect disequilibrium abundances of {\cotwo} in brown dwarf with higher metallicity. As our models have solar metallicity, we do not specifically discuss these signatures here.

\begin{figure*}
  \centering
  \includegraphics[width=0.9\textwidth]{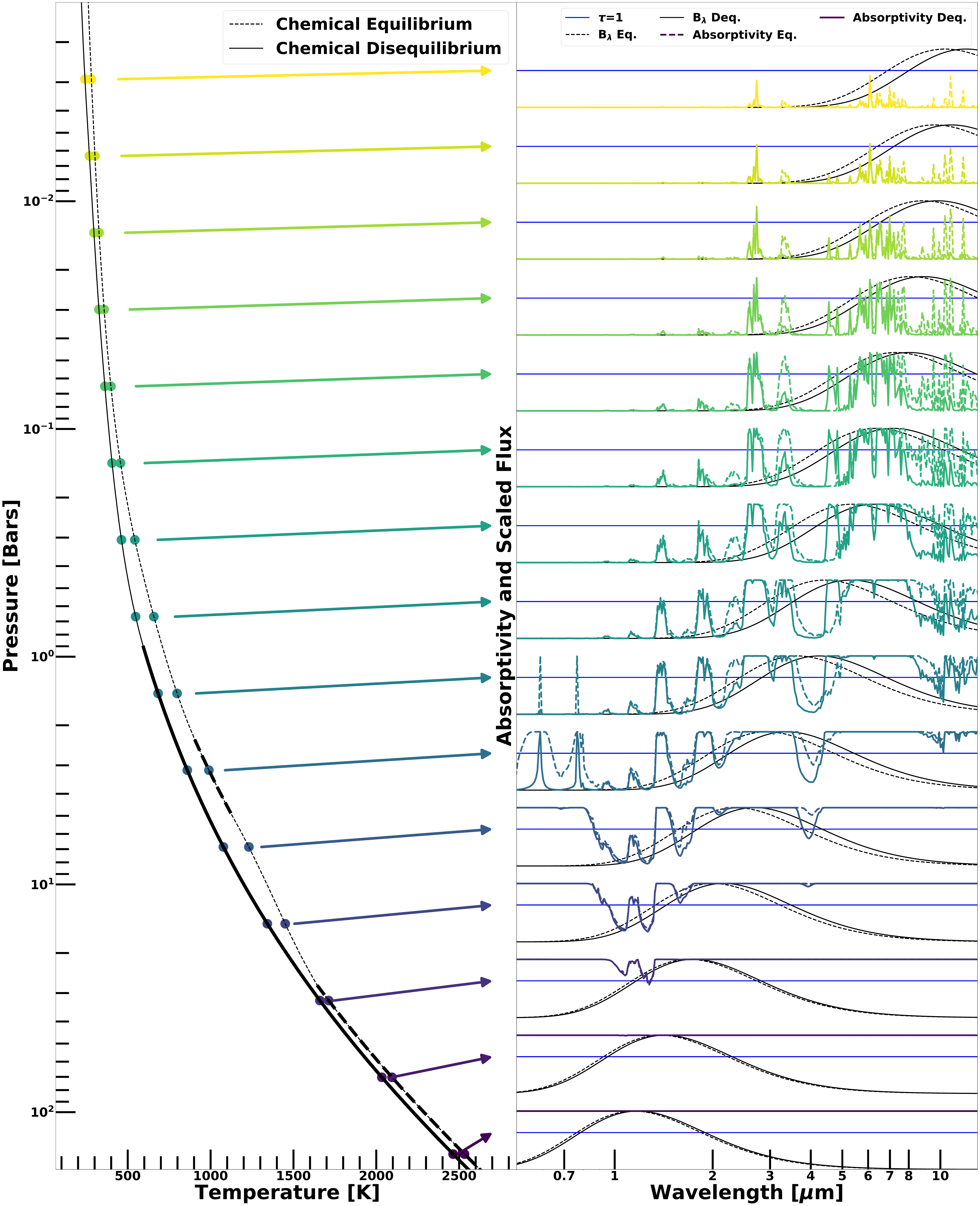}
  \caption{{\bf Left panel} shows the $T(P)$ profile of a brown dwarf with {\teff} of 700 K and log(g) of 4.5. The dashed line is the $T(P)$ profile when chemical equilibrium is assumed whereas the solid line shows the $T(P)$ profile for disequilibrium chemistry. The convective zones are shown with darker and thicker lines. {\bf Right panel} shows the wavelength dependant absorptivity of the different atmospheric layers for both the cases. The absorptivity is given by 1-e$^{-\tau}$. The absorptivity for the equilibrium chemistry model is shown with dashed lines whereas that for disequilibrium chemistry model are shown with solid lines. The local normalized blackbody spectrum is also shown for each layer with dashed and solid lines for equilibrium and disequilibrium chemistry models, respectively. The blue horizontal line for each layer shows the absorptivity corresponding to ${\tau}$ =1. {\bf Main Point} -- Opacity changes lead to large ($\sim$100 K) differences in $T(P)$ profiles between chemical equilibrium and disequilibrium models.}
\label{fig:figconrad}
\end{figure*}

Of particular interest to our work here are the changes in $T(P)$ structure, and the locations of radiative and convective zones show in Figure \ref{fig:fig2}. We examine the causes for these differences in more detail in Figure \ref{fig:figconrad} for a brown dwarf with {\teff} = 700 K and log($g$) of 4.5. The left panel in Figure \ref{fig:figconrad} shows the $T(P)$ profile of this brown dwarf under equilibrium chemistry with the dashed line and the $T(P)$ profile of the brown dwarf with chemical disequilibrium is shown with the solid line. The radiative zones are shown with narrower lines whereas the convective zones are shown with thicker lines.

There are significant changes in the location and number of radiative and convective zones in the two profiles, with the disequilibrium model having one convective zone at depth and radiative zone at lower pressure.  However, the equilibrium models finds two convective zones and two radiative zones. To examine why, the right panel in Figure \ref{fig:figconrad} shows the wavelength dependent absorptivity in several layers of these atmospheres with dashed (equilibrium chemistry) and solid (disequilibrium chemistry) lines. The absorptivity is given by A=(1-e$^{-\rm \tau}$). A high absorptivity (close to 1) corresponds to an opaque wavelength whereas a low absorptivity corresponds to a transparent wavelength window. The normalized local blackbody Planck function B$_{\lambda}$ is also plotted for each layer for the equilibrium (dashed black line) and disequilibrium chemistry (solid black line) models for comparison with the absorptivity. We follow \citet{marley15} for the explanation of this phenomenon below. 

As the atmosphere becomes colder with lower pressure in both models, the peak of the local B$_{\lambda}$ shifts to longer wavelengths. In the deep atmosphere $\sim 100$ bar, gaseous optical depths are very high which causes the absorptivity to be very close to 1 across all wavelength ranges for both cases. The atmosphere is opaque to radiative energy transport in these pressure layers for both cases, and convection ensues at these pressure levels. This is the case at pressures higher than $\sim$ 40 bars for both models. At about 20 bars some transparent windows start to appear. Since the equilibrium chemistry model is hotter than the disequilibrium chemistry model, the peak of its local blackbody emission coincides more closely in wavelength with the transparent opacity windows between 0.8-1 $\mu$m in the atmosphere at these pressures. This causes the equilibrium chemistry model to be radiative in these pressure ranges whereas the disequilibrium chemistry model is still convective as its local blackbody emission still peaks at longer wavelengths where the atmosphere is opaque at $\sim$ 20 bars pressure. As the atmosphere gets colder with lower pressure, large transparent windows develop between 0.5-2 $\mu$m but as these layers are much colder, their blackbody peaks are at wavelengths greater than 2 $\mu$m.

At $\sim$ 2 bars, the blackbody peak of the chemical equilibrium model hits another opaque range between 2-3 $\mu$m. This causes it to develop another, detached, convective zone at these pressures. At pressures lower than this range, a large transparent opacity window develops between 3.5-5 $\mu$m in both the models.  The disequilibrium model is more transparent than the equilibrium model between 3.5-4.5 $\mu$m (due to less \meth\ and \water) but is more opaque than the equilibrium model between 4.5-5 $\mu$m (due to more \co). As the local blackbody function of both models also peaks between 3.5-5 $\mu$m in these pressure layers, these differences in transparency between the two atmospheres causes a large change ($\sim$ 150 K) between their $T(P)$ profiles in these pressures.  As the cooler local blackbody peak of the disequilibrium model more closely overlaps with the opaque window in the disequilibrium model between 4.5-5 $\mu$m, its atmosphere is still convective in these pressures, while the equilibrium model is radiative. At pressures lower than $\sim$ 0.8 bars, both atmospheres are radiative as more transparent windows start appearing in longer wavelengths with decreasing pressure. With these differences in the $T(P)$ profile and convective/radiative zones between models with chemical equilibrium and disequilibrium, the next important question to be investigated is whether gases like {\co}, {\amon}, and {\water} are quenched in the radiative or convective zones of models with chemical disequilibrium. Results from our model grid on this question are presented next. 

\subsection{Quenching in the Radiative Zone or the Convective Zone ?}\label{sec:rad_conv}

\begin{figure*}
  \centering
  \includegraphics[width=1\textwidth]{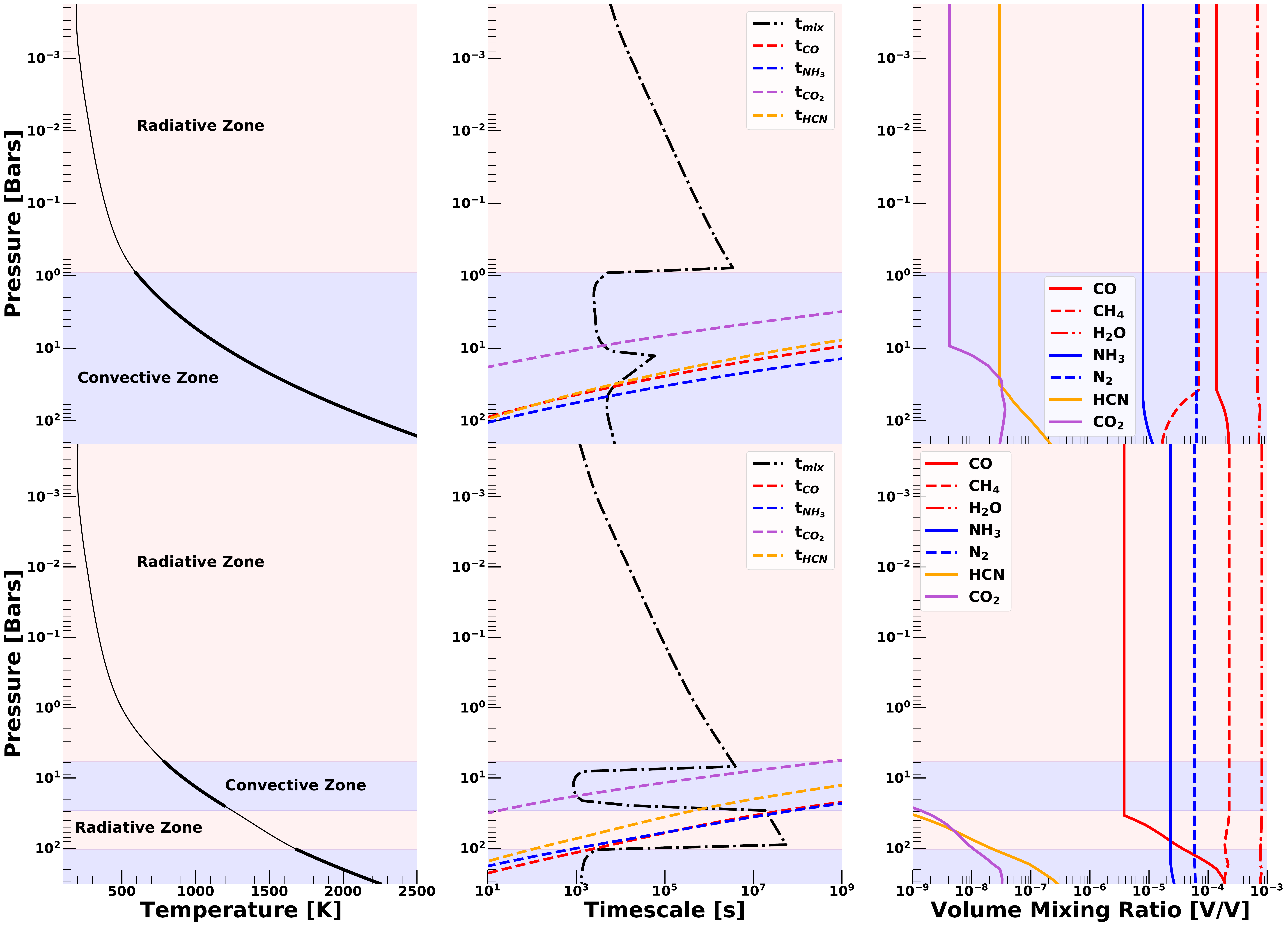}

\caption{Model $T(P)$ profile, chemical and dynamical timescales, and chemistry for a brown dwarf with \teff\ 700 K and log(g) = 4.5 is shown in the top panels whereas the same properties for a brown dwarf with \teff\ 700 K and log(g) = 5.25 is shown in the bottom panels. $T(P)$ profile of the two brown dwarfs are shown in the {\bf left panels} -- one with a single deep convective zone while another with a deep ''sandwiched" radiative zone between two deep convective zones. The {\bf middle panels} show the comparison between the mixing timescale (black) and the chemical timescales of different gases like CO and NH$_3$ as a function of pressure for each of the $T(P)$ profiles shown in the left columns. The {\bf right panels} show the volume mixing ratios of various molecules as a function of pressure for each atmosphere. {\bf Main Point} -- Gases can be quenched in a deeper ``sandwiched" radiative zone in some brown dwarfs, where mixing timescales can be orders of magnitude larger than in convective zones.} 
\label{fig:fig3}
\end{figure*}

Quenching of gases can occur in convective zones or radiative zones. Vertical mixing in convective zones is expected to be very efficient, yielding a high {\kzz} and correspondingly low {\tmix}. The top three panels of Figure \ref{fig:fig3} show the $T(P)$ profile of a brown dwarf with \teff\ of 700 K and log(g) of 4.5 (left panel), the mixing and chemical timescales ({\tmix} and {\tchem}) for the same brown dwarf (middle panel), and the volume mixing ratios of various molecules (right panel) in the brown dwarf atmosphere. This brown dwarf has a single convective and radiative zone. The mixing timescale {\tmix} is shown in dashed black whereas the {\tchem} for \co\ and other chemical reactions are shown with different colors of dashed lines. The {\tmix} profile has been calculated using mixing length theory in the convective zone and with Equation \ref{eq:kzz_rad} in the radiative zone. The mixing timescales in radiative zones are expected to be longer at depth, due to the lack of efficient dynamical mixing processes. However, as briefly described in \S\ref{sec:model}, the mixing timescale is expected to be short in the much lower pressure regions (P $\sim$ 10$^{-4}$ bars) due to the presence of gravity wave breaking mechanisms \citep{freytag10,moses05,moses11}.

{\tmix} is much longer than {\tchem} at pressures $>100$ bars and therefore {\co} and {\meth} are expected to follow chemical equilibrium in these high pressure high temperature parts of the atmosphere. But the {\tchem} for CO$\leftrightarrow$CH$_4$ reaction is a very strong function of both temperature and pressure \citep{Zahnle14}. Therefore, it increases by 12 orders of magnitudes within one order of magnitude in pressure and a change of 500 K in temperature. As a result, the two timescales cross each other at about $\sim$ 30 bars at the quench pressure ($P_{\rm Q}$). The abundances of both of these molecules, and {\water}, are quenched to constant values above $P_{Q}$ as a result, as shown in the Figure \ref{fig:fig3} top right panel. As different net chemical reactions proceed at different rates and have \emph{different} {\tchem}, their {\pq} will also be different.

Self-consistent chemical equilibrium models of T-dwarfs with {\teff} within the range of 400--800 K have shown the presence of an upper detached convective zone in addition to the deeper convective zone \citep{burrows97,marley21}, with a radiative zone in between. The $T(P)$ profile in this scenario is shown in Figure \ref{fig:fig3} bottom left panel for a brown dwarf with {\teff} of 700 K (as above) but with a log(g) of 5.25. This detached zone appears when the $T(P)$ profile of the atmosphere is such that the radiative energy transport in the deeper atmosphere becomes efficient due to the opening of a molecular opacity window at near infrared wavelengths \citep{marley15}, as shown in Figure \ref{fig:fig3}. The resulting ``sandwiched radiative zone'' is very likely to have a much slower rate of vertical mixing (longer {\tmix}) compared to the convective zones. As these radiative zones appear at high pressures (P $\ge$ 10 bars), it is possible that molecules like {\co}, {\meth} and {\amon} are quenched in the radiative zone instead of the convective zone. Note that the mixing timescale profile in the upper middle panel in Figure \ref{fig:fig3} shows a ``kink" like structure in {\tmix} values between 10-30 bars. This kink is due to the decrease in the convective heat flux ($F$ in Equation \ref{eq:kzz_con}) in the convective zone around that region. This can happen in convective regions where a non-negligible, but smaller than the convective heat transport, amount of energy transfer in a convective zone still occurs via radiation. This leads to a smaller convective heat flux in these regions.

The bottom middle panel of Figure \ref{fig:fig3}  shows this scenario. Within our modeling framework, the {\tmix} is low (high {\kzz}) in the deep convective zone and {\tmix} is high (low {\kzz}) in the deep sandwiched radiative zone. The {\pq} for {\co} in this case is within the sandwiched radiative zone. Therefore, if one is trying to probe the {\kzz} of this hypothetical object with constant {\kzz} models \citep[e.g.][]{Miles20}, one would infer a much lower {\kzz} than what is expected from Equation \ref{eq:kzz_con}. However, the reason for this is is not because of sluggish mixing in the convective zone but rather quenching is occurring in a radiative zone where mixing is expected to be sluggish.

Such a scenario was suggested in fits to observations presented in \citet{Miles20}. They inferred the {\kzz} for several late T and early Y-dwarfs and found that the {\kzz} for objects with {\teff} between 400-800 K are much smaller than expectations from mixing length theory. Although equilibrium chemistry models find the sandwiched radiative zone between {\teff} of 400--800 K \citep[e.g.][]{marley21,allard2012}, it is not clear if the self-consistent treatment of disequilibrium chemistry preserves these sandwiched radiative zones and also if molecules like {\co} or {\amon} actually quench in these sandwiched radiative zones. Next, we investigate this scenario in more detail.

\begin{figure}
  \centering
  \includegraphics[width=0.45\textwidth]{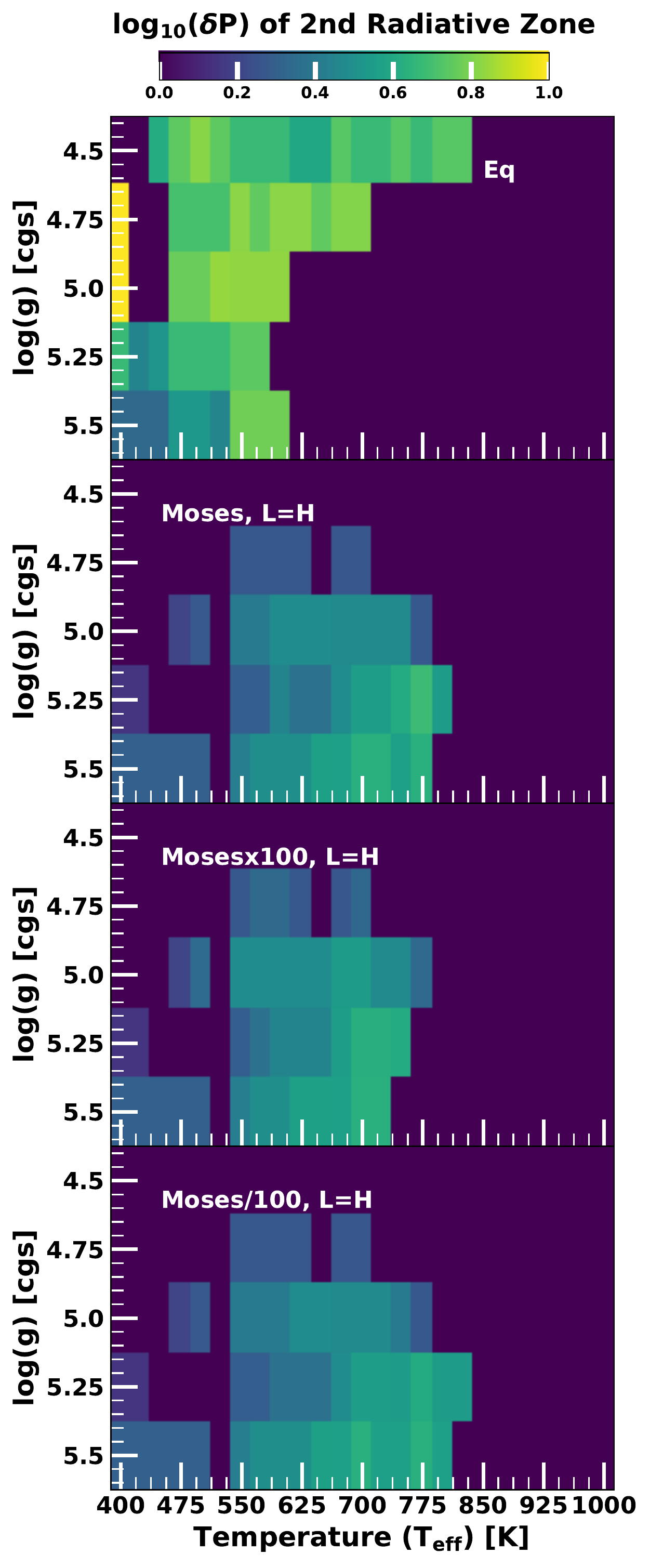}

  \caption{Appearance of the 2nd radiative zone in the {\teff} vs. log(g) parameter space in our grid is shown. The colors show the value of the quantity log$_{10}$(P$_{\rm bot}$/P$_{\rm top}$) where P$_{\rm bot}$ and P$_{\rm top}$ are the pressures at the bottom and top of the sandwiched radiative zone, respectively. The {\bf top panel} shows this in case of equilibrium chemistry. The subsequent panels show the presence of the 2nd radiative zone if different vigor of {\kzz} is chosen keeping the convective zone mixing length fixed at $H$. Dark blue indicates no second convective zone at depth.}
\label{fig:fig4}
\end{figure}

\begin{figure}
  \centering
  \includegraphics[width=0.45\textwidth]{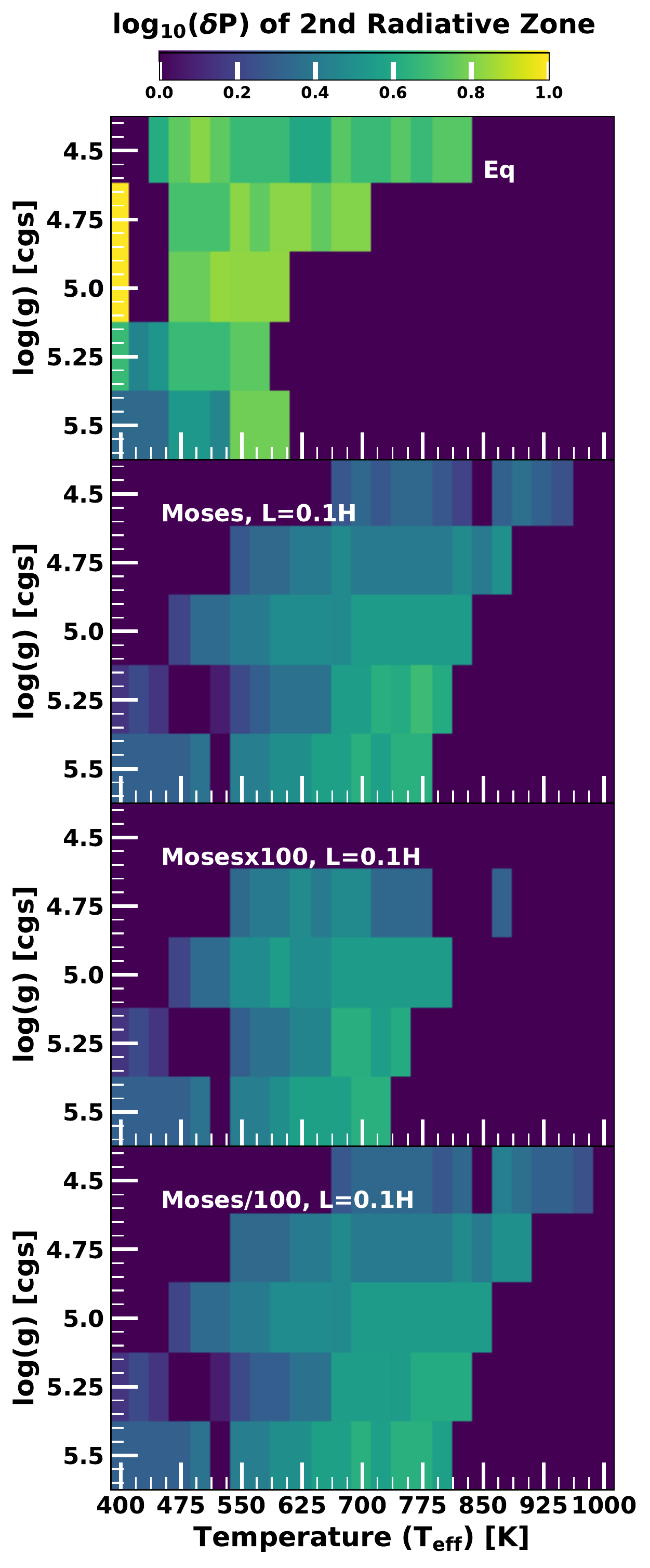}

  \caption{Appearance of the 2nd radiative zone in the {\teff} vs. log(g) parameter space in our grid is shown. The colors show the value of the quantity log$_{10}$(P$_{\rm bot}$/P$_{\rm top}$) where P$_{\rm bot}$ and P$_{\rm top}$ are the pressures at the bottom and top of the sandwiched radiative zone, respectively. The {\bf top panel} shows this in case of equilibrium chemistry. The subsequent panels show the presence of the 2nd radiative zone if different vigor of {\kzz} is chosen keeping the convective zone mixing length fixed at $0.1\times{H}$. Dark blue indicates no second convective zone at depth.}
\label{fig:fig4con}
\end{figure}

We can expand the modeling work shown in Figure \ref{fig:fig3} over a range in \teff\ and log (g) to assess under what conditions a sandwiched radiative zone occurs in self-consistent non-equilibrium models. Figure \ref{fig:fig4} shows the appearance of the sandwiched radiative zones across this phase space in the presence and absence of vertical mixing. The colors shown in Figure \ref{fig:fig4} represents log$_{10}$(P$_{\rm bot}$/P$_{\rm top}$) where P$_{\rm bot}$ and P$_{\rm top}$ are the pressures at the bottom and top of the sandwiched radiative zone, respectively. If the model does not find a sandwiched radiative zone, this number will be zero. The top panel in Figure \ref{fig:fig4} shows the sandwiched radiative zones in chemical equilibrium models.  The other three panels show the appearance of the second radiative zone with three different magnitudes of mixing in the radiative zone, all assuming the mixing length $L$ is $H$ in the convective region.  Under the assumption of equilibrium chemistry it is clear that the sandwiched radiative zones appear between {\teff} of 450-800 K for a log(g) of 4.5. However, for objects with log(g) of 4.75 and higher, the sandwiched radiative zones only appear between 450-600 K in chemical equilibrium models. For models with disequilibrium chemistry, the models that show the presence of this sandwiched radiative zone appear at a different set of {\teff} and log(g) values. As can be seen in Figure \ref{fig:fig4} (second, third, and fourth panels), high gravity models with log(g) of 5.0 and greater have this sandwiched radiative zone between {\teff} of $\sim$ 500-800 K. For models with log(g) of 4.75 the sandwiched radiative zone appears between 500-750 K. However, for models with log(g) of 4.5, the sandwiched radiative zone is non-existent when quenching due to vertical mixing is taken into account.  The thickness of these zones is typically smaller, too, as can be seen from the shading that is less yellow.

Figure \ref{fig:fig4con} shows the appearance of this sandwiched radiative zone if the convective mixing length $L$ is smaller, only  $0.1\times{H}$. The second, third, and fourth panels in Figure \ref{fig:fig4con} show that larger number of models now develop this sandwiched radiative zone compared to models with $L=H$ shown in Figure \ref{fig:fig4}. The sandwiched radiative zone does indeed appear at low gravity (log(g) = 4.5) when $L= 0.1\times H$.  Additionally, the radiative zones tend to appear at higher \teff\ than for $L=H$ models.  This shows that the strength of vertical mixing in both the radiative and convective zone can influence the presence and absence of multiple convective zones.  

\begin{figure}
  \centering
  \includegraphics[width=0.45\textwidth]{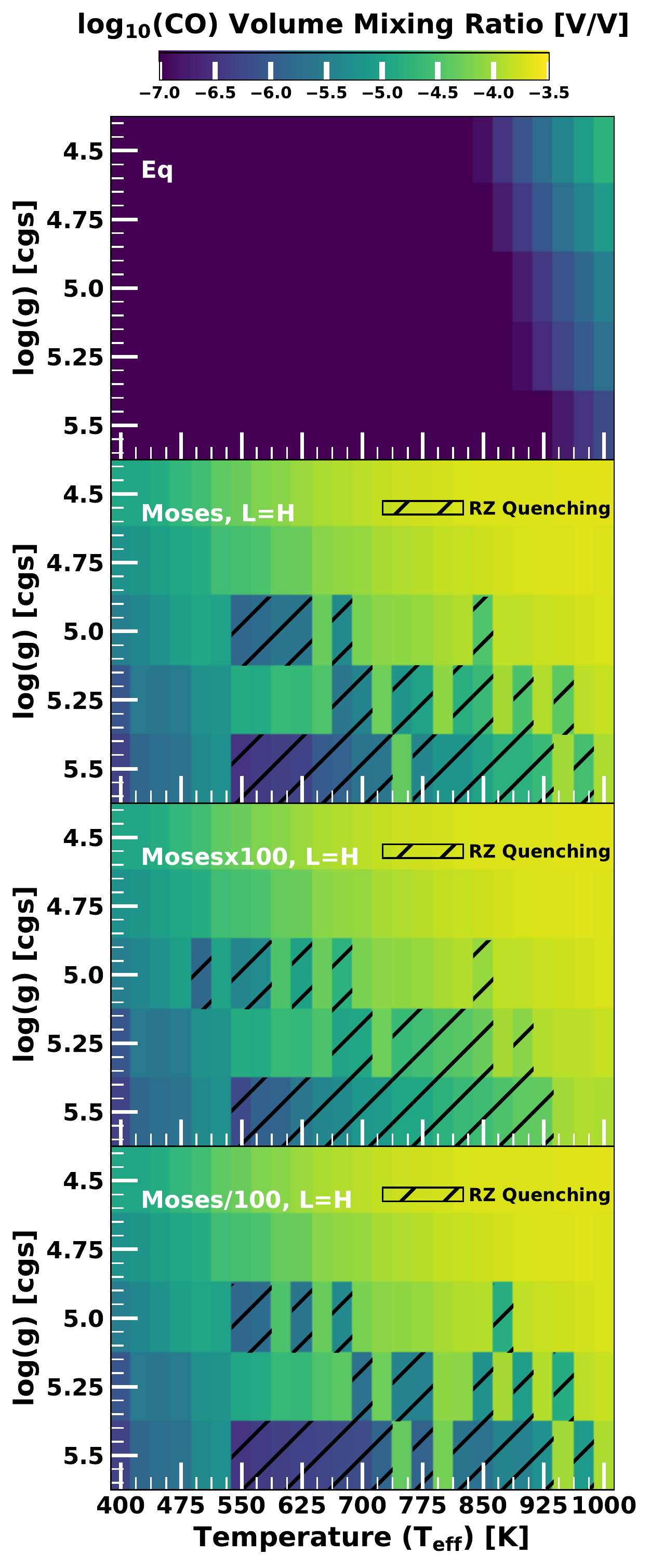}
  \caption{ {\bf Top panel} shows the equilibrium chemistry abundance of \co\ at pressures where $T$= \teff\ in our \teff\ vs. gravity grid of chemical equilibrium models. The {\bf second, third and last panels} depict the quenched \co\ abundance within the \teff\ vs. gravity parameter space in our disequilibrium chemistry models. The second, third and last panels correspond models with radiative zone \kzz\ values of ``Moses", ``Moses$\times$100", and ``Moses/100", respectively while the convective zone mixing length L was fixed at the scale height H. Models where \co\ was quenched at the radiative zone are marked with black hatches. \co\ was quenched at a convective zone in the rest of the models. Models with radiative zone \co\ quenching show much smaller \co\ abundance compared to models with convective zone quenching.}
\label{fig:fig5}
\end{figure}

The effect of radiative and convective \kzz\ on the chemical abundances is  more complex than it may appear.  The presence of a sandwiched radiative zone  does not guarantee that quenching occurs \emph{in} the radiative zone. Furthermore, radiative zone quenching can also occur in atmospheres with a single radiative zone and single convective zones. We investigate the resulting photospheric {\co} abundances in \teff\ vs. log(g) space in Figure \ref{fig:fig5}.  The top panel shows the {\co} abundance in models in chemical equilibrium. The second, third, and fourth panels show models with disequilibrium chemistry. Each of these panels represents models with different values of {\kzz} in the radiative zone while the mixing length in the convective zone is fixed to $H$. Models where {\co}, {\meth}, and \water\ are quenched in radiative zones are marked with black hatches. The rest of the models still have {\co} quenching, but in the convective zones instead. 

The \co\ abundances in these models with radiative zone \co\ quenching also show large differences from \co\ abundances in models with convective zone quenching. These differences appear as discontinuities in the \co\ abundance heat map shown in the second, third, and fourth panels of Figure \ref{fig:fig5}. The reason behind these discontinuities is mainly the smaller \pq\ of \co\ in models with radiative zone quenching compared to models with convective zone quenching.

\begin{figure}
  \centering
  \includegraphics[width=0.45\textwidth]{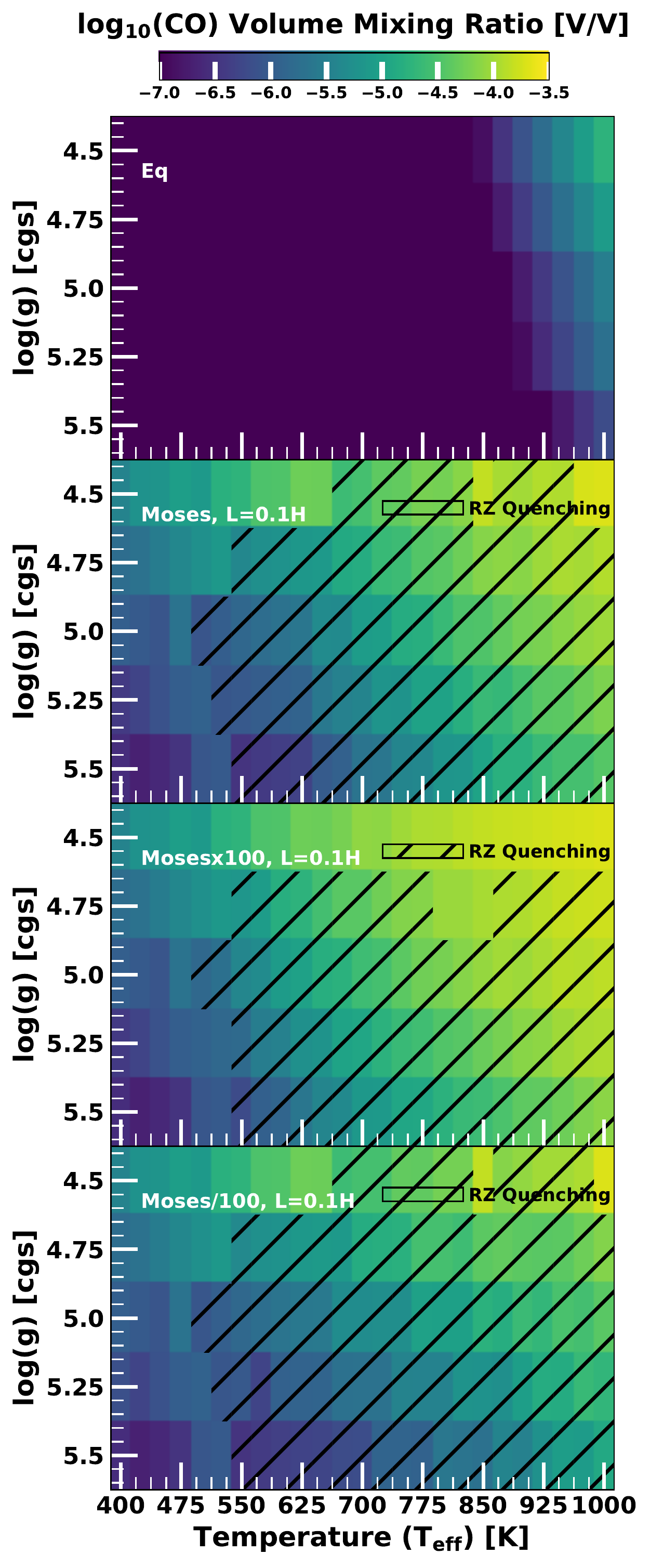}
  \caption{Same as Figure \ref{fig:fig5} except with the mixing length in convective zones fixed at 0.1$\times$H. In these models \co\ is mostly quenched in a radiative zone.}
\label{fig:fig6}
\end{figure}

However, Figure \ref{fig:fig6} shows that \co\ abundances over \teff\ and log(g) will change if the mixing length in the convective zone is smaller.  Here, again, $L = 0.1\times H$, leading to smaller {\kzz} values, and longer {\tmix} in the convective zone. This causes quenching to often occur in the top or sandwiched radiative zone in a significantly larger fraction of the {\teff} and log(g) parameter space compared to the nominal $L=H$ scenario in Figure \ref{fig:fig5}. For models with log(g)=4.75 and greater, all models with {\teff} $\ge$ 550 K show radiative zone quenching of {\co} and {\meth} in this case. Both Figure \ref{fig:fig5} and \ref{fig:fig6} makes it clear that the chemical properties of brown dwarfs that will have either radiative zone or convective zone quenching depends on the vigor of mixing in both of these types of zones.

\begin{figure}
  \centering
  \includegraphics[width=0.45\textwidth]{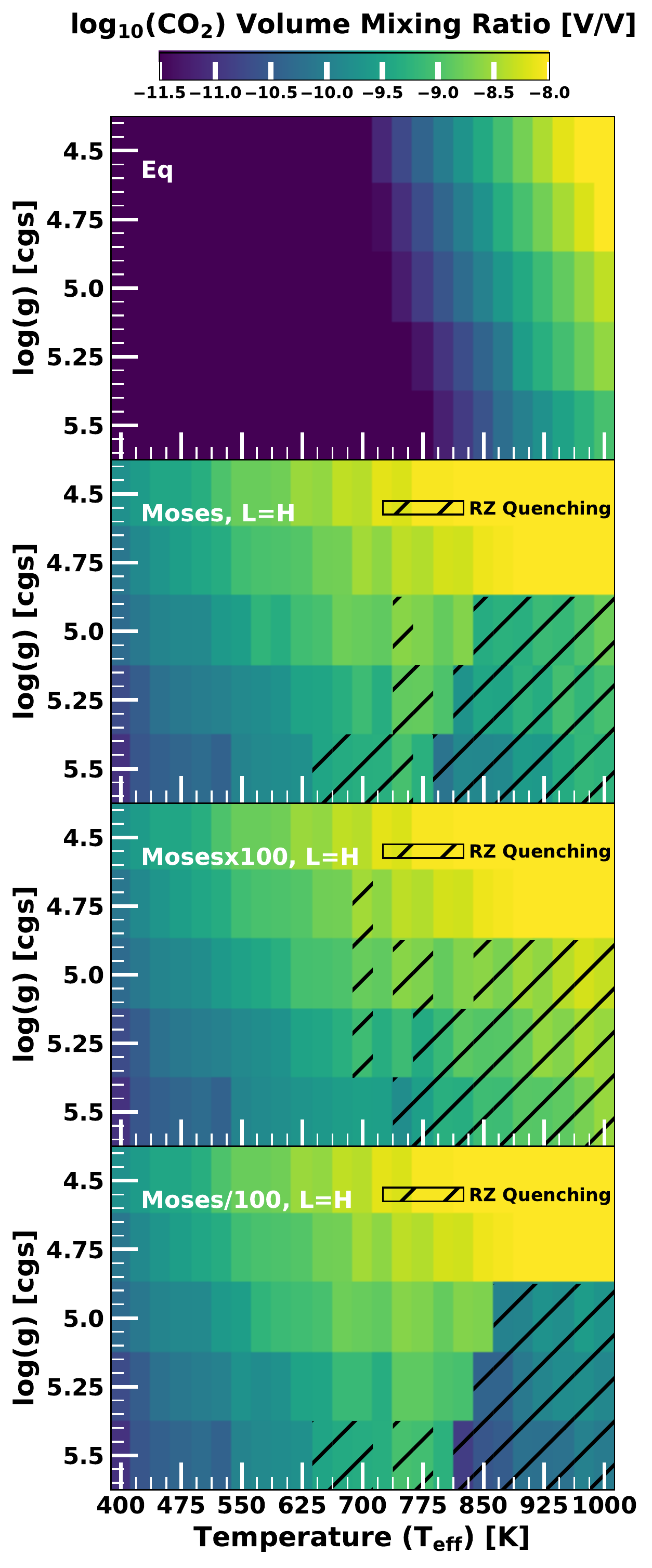}
  \caption{Same as Figure \ref{fig:fig5} but for quenching of {\cotwo}. Given the different \tchem\ for \cotwo, compared to \co, the quench levels are entirely different.}
\label{fig:fig7}
\end{figure}

\begin{figure}
  \centering
  \includegraphics[width=0.45\textwidth]{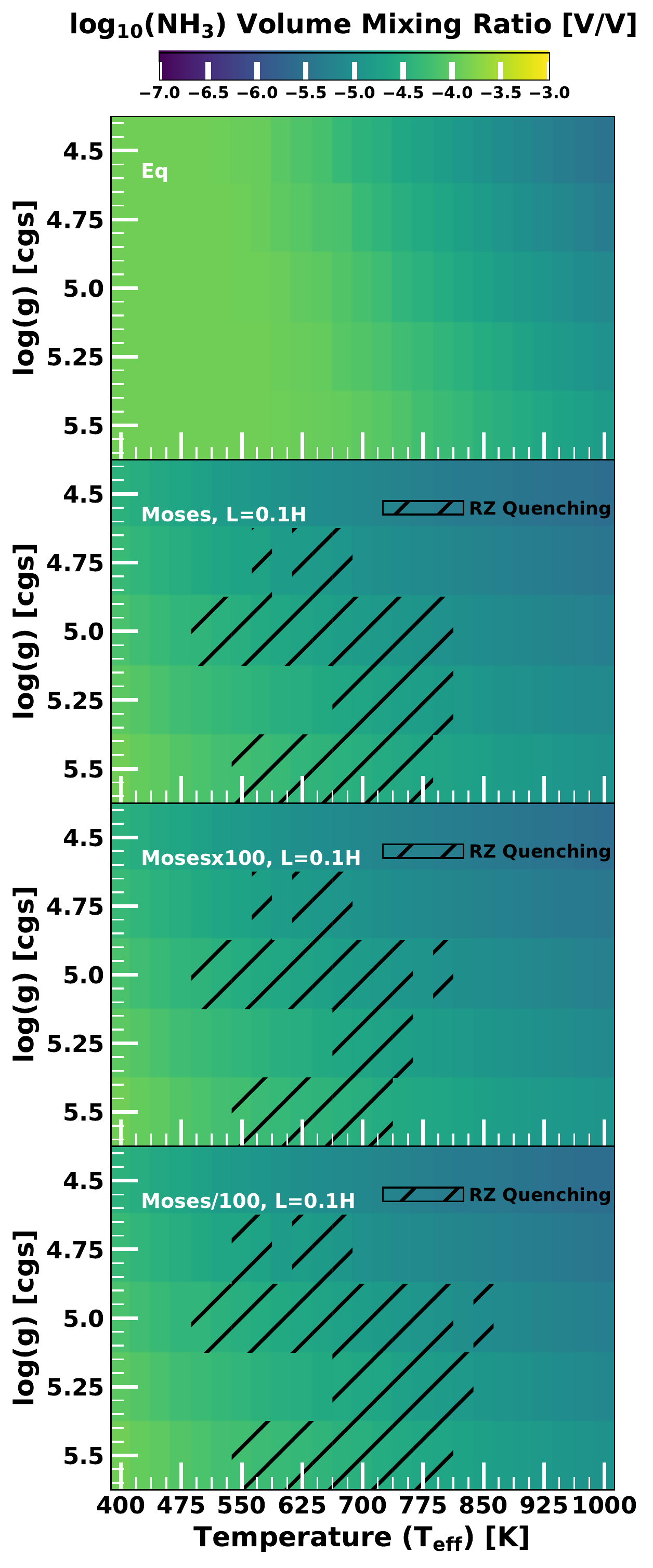}
  \caption{Same as Figure \ref{fig:fig6} but for quenching of {\amon}. Given the different \tchem\ for {\amon}, compared to \co, the quench levels are entirely different.}
\label{fig:fig6p}
\end{figure}

Furthermore, since {\tchem}, and hence the {\pq}, of different gaseous species are different, this means that models where {\co} quenches in radiative zones are not necessarily models where {\cotwo} or {\amon} also quenches in the radiative zone. This effect can be clearly seen in Figure \ref{fig:fig7}, which shows the same set of models in Figure \ref{fig:fig5}, but now with the {\cotwo} abundances and its radiative/convective quench behavior. Because the {\tchem} for the thermochemical equilibrium of {\cotwo} is orders of magnitude shorter than for {\co} and {\meth}, the models where {\cotwo} quenches in the radiative zone are also different than those where {\co}/{\meth} quenches in the radiative zone. If the mixing length in the convective zone is $H$, \cotwo\ radiative zone quenching occurs mostly in models that are hotter than 700 K and have gravity higher than log(g) of 5. The same effect for quenching of {\amon} has been shown in Figure \ref{fig:fig6p} in models with $L=0.1\times{H}$. Comparing Figure \ref{fig:fig6p} with \ref{fig:fig6} shows that the radiative zone quenching of \amon\ occurs in only a small subset of lower temperature (500--800 K) models with radiative zone quenching of \co\ , when $L=0.1\times{H}$. The \tchem\ for the reaction between N$_2\leftrightarrow${\amon} is the slowest among the other reactions described in \S\ref{sec:deq_py}. As a result, if $L=H$, \amon\ always gets quenched in the deep convective zone in the parameter space explored by our grid. Even if the convective mixing length is smaller ($L=0.1\times{H}$), \amon\ gets quenched in the convective zone of the majority of the models except in some where deeper ``sandwiched" radiative zones develop. But in order to measure these quenched abundances observationally and infer the {\kzz} profile of the atmosphere, we need to investigate the effect of \kzz\ on observables, such as the emission spectra of these brown dwarfs, which we discuss next.

\subsection{Effect of Radiative and Convective Zone {\kzz} on Emission Spectra}\label{sec:spectrum}

\begin{figure*}
  \centering
  \includegraphics[width=0.9\textwidth]{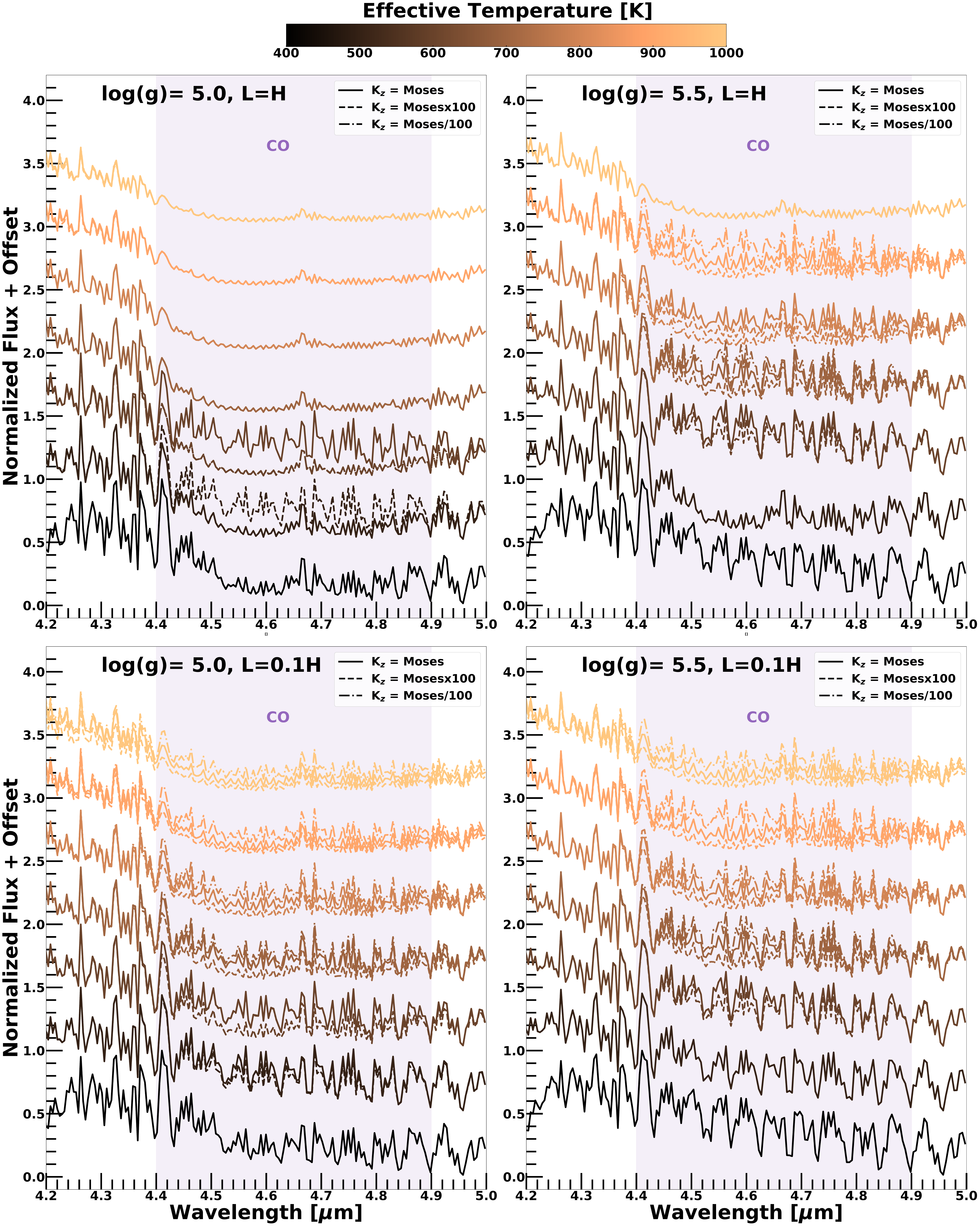}
  \caption{M-band Thermal emission spectra across a range of \teff\ between 400 K and 1000 K with steps of 100 K are shown in each panel. log ($g$) = 5.0 is on the left and 5.5 is on the right. The colors are representative of the {\teff}. The models shown in the upper left and right panels assume the convective mixing length $L$ to be $H$ whereas the models shown in the lower left and right panels assume $L=0.1\times{H}$. The three line styles show different radiative zone {\kzz} parametrizations, including Moses (solid), Moses/100 (dashed), and Moses$\times$100 (dash-dot).  If quenching occurs in a convective zone the three spectra plot atop each other. The shaded region shows the wavelength range where {\co} is the most important absorber.}
\label{fig:fig8}
\end{figure*}

\begin{figure*}
  \centering
  \includegraphics[width=0.9\textwidth]{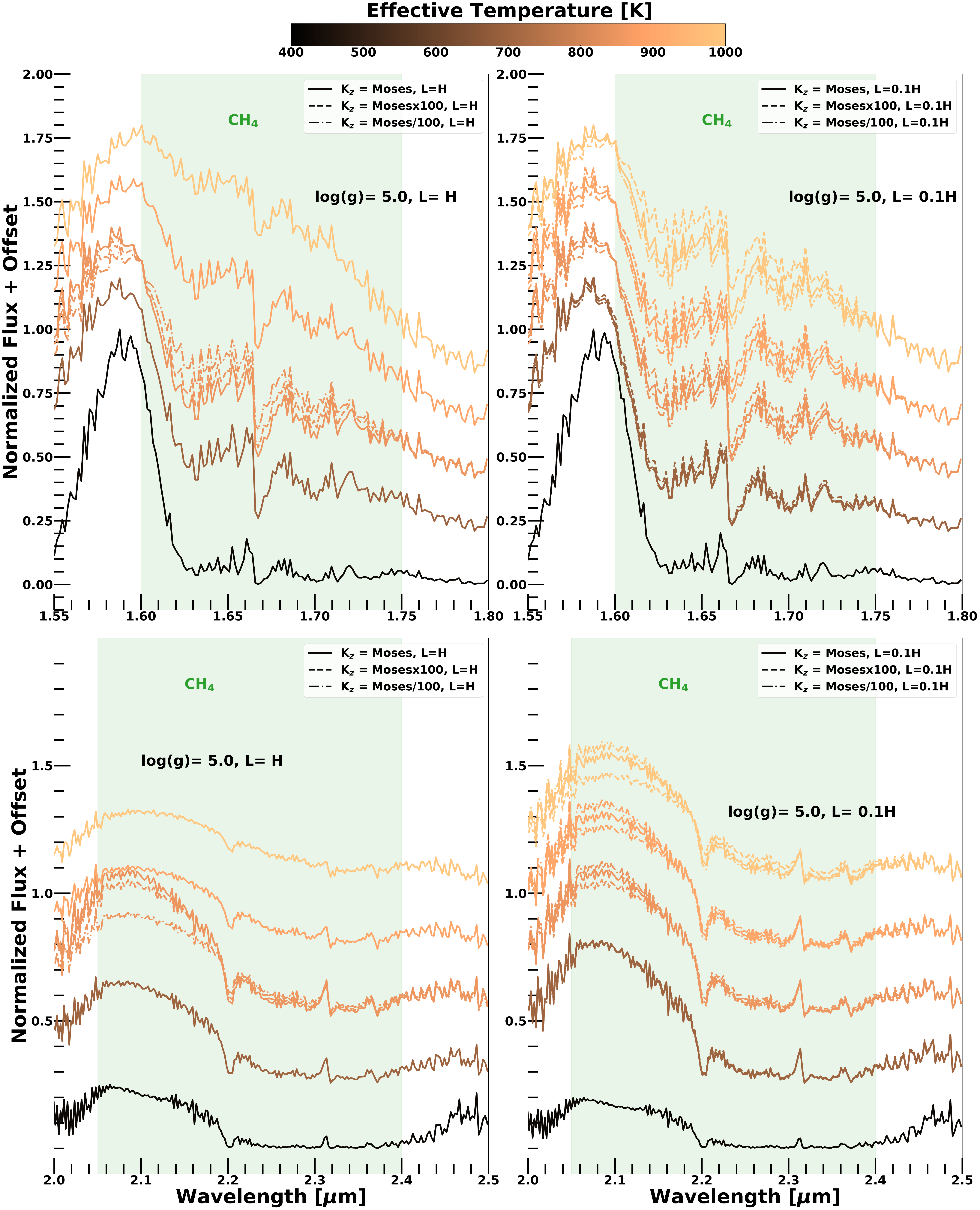}
  \caption{H-band thermal emission spectra of models from our grid are shown in the top left and right panels whereas the bottom left and right panels show the K-band emission spectra. Models with log(g)=5 and different {\teff} between 400-1000 K are shown in each panel. From bottom to top, spectra for \teff\ values of 425 K, 700 K, 850 K, 900 K, and 1000 K are shown in all the panels. {\bf Top left panel} shows the thermal emission spectra of models which have convective mixing length set at $H$ but have different {\kzz} parametrizations like Moses, Moses/100, and Moses$\times$100 which are represented by different line types in the panels like solid, dashed, and dashed-dotted. {\bf Top right panel} shows the same but for models with convective mixing length set at $0.1\times{H}$. {\bf Bottom left and right panels} show the sensitivity of the K band spectrum to radiative zone \kzz\ when $L=H$ and $L=0.1\times{H}$, respectively.  The shaded region shows the wavelength range where {\meth} is the most important absorber.}
\label{fig:fig1pt5to1pt8}
\end{figure*}

\begin{figure*}
  \centering
  \includegraphics[width=0.9\textwidth]{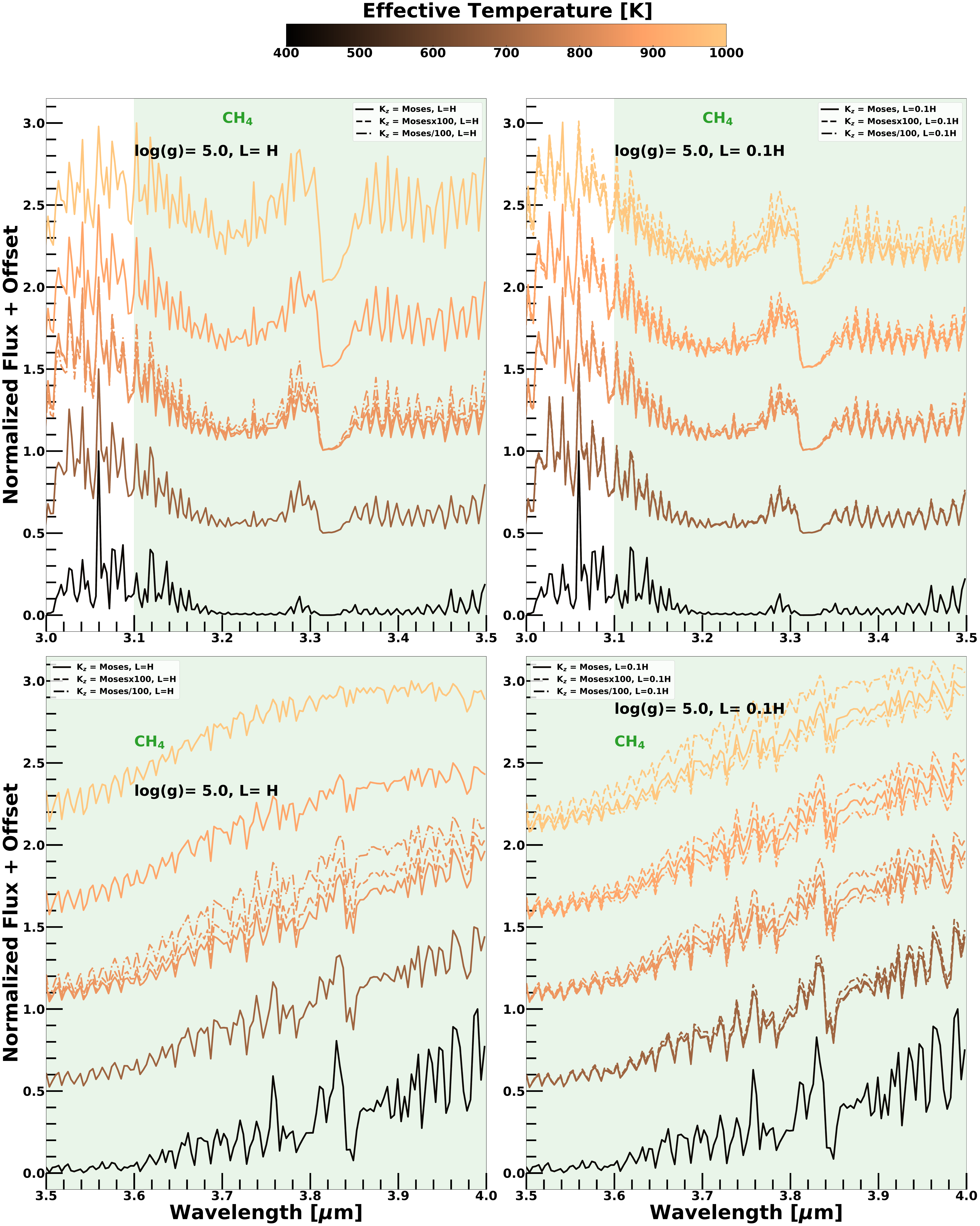}
  \caption{Thermal emission spectra of brown dwarfs between 3 and 3.5 $\mu$m are shown in the top left and right panels. Emission spectra of the same models in the L-band are shown in the bottom left and right panels. Models with log(g)=5 and different {\teff} between 400-1000 K are shown in each panel. From bottom to top, spectra for \teff\ values of 425 K, 700 K, 850 K, 900 K, and 1000 K are shown in all the panels. {\bf Top left panel} shows the emission spectra of models which have convective mixing length set at $H$ but have different {\kzz} parametrizations like Moses, Moses/100, and Moses$\times$100 which are represented by different line types in the panels like solid, dashed, and dashed-dotted. {\bf Top right panel} shows the same but for models with convective mixing length set at $0.1\times{H}$. {\bf Bottom left and right panels} show the sensitivity of the L-band spectrum to radiative zone \kzz\ when $L=H$ and $L=0.1\times{H}$, respectively. The shaded region shows the wavelength range where {\meth} is the most important absorber.} 
\label{fig:fig2to2pt5}
\end{figure*}

We have established in \S\ref{sec:rad_conv} that gases like {\co}, {\meth} and {\cotwo} can be quenched in the sandwiched radiative zones or single radiative zones in a subset of late T and early L-dwarfs. In Figures \ref{fig:fig8}, \ref{fig:fig1pt5to1pt8}, and \ref{fig:fig2to2pt5},   we show how thermal emission spectra are sensitive to the vigor of vertical mixing in the radiative zone of these objects. These spectra, with a spectral resolution of 600, are calculated using the thermal emission calculation module of \texttt{PICASO} \citep{batalha19}, using the same opacities that were used for model convergences to RCE. Sources of these opacities can be found in \citet{marley21}.

The thermal emission spectra between 4.2-5.0 $\mu$m, across a range of \teff\ between 400 K and 1000 K with steps of 100 K, are shown for log(g) values of 5 and 5.5 in the two upper panels of Figure \ref{fig:fig8}. The choice of the wavelength range focuses on the role of {\co}, which has significant impacts on the spectrum in M-band between 4.4-5 $\mu$m. \water\ is the primary absorber between 4.0-4.4 $\mu$m. These models assume $L=H$ in the convective zones. Figure \ref{fig:fig8} lower panels shows the spectrum from the same set of models except $L= 0.1\times{H}$. Spectra arising from different vigors of vertical mixing in the radiative zones are shown with different line dashes in all the panels. 


Sensitivity of the M-band spectrum to radiative zone {\kzz} values is more common for higher gravity objects with log(g) of 5.0 and 5.5. If the convective mixing length is $H$, then this sensitivity appears between models with {\teff} 550- 700 K for log(g) of 5. However, if the convective mixing length is $0.1\times{H}$, then this sensitivity appears much more commonly between models with {\teff} of 500-1000 K for log(g)= 5.0. At higher gravity, log(g) of 5.5, the sensitivity of M-band spectra to radiative zone {\kzz} appears at both convective mixing lengths of $H$ and $0.1\times{H}$. In Figure \ref{fig:fig8} upper right panel, this sensitivity appears between 500-900 K whereas in Figure \ref{fig:fig8} lower right panel this sensitivity appears between 500-1000 K.

While the M-band shows the greatest sensitivity to \kzz\ in the radiative and convective zones, smaller differences in the spectra arise in other wavelengths as well. Figure \ref{fig:fig1pt5to1pt8} top panels shows the sensitivity of the H-band spectra to the radiative zone \kzz\ across various values of \teff\ for brown dwarf models with log(g)=5.0. The top left panel shows models with $L=H$ whereas the top right panel shows models with $L=0.1\times{H}$. The sensitivity mainly appears between 1.69 $\mu$m and 1.75 $\mu$m in models where \meth\ is quenched in the radiative zone as it is the main opacity source at these wavelengths. \meth\ is also the main opacity source between 2--2.2 $\mu$m, 3.2-3.5 $\mu$m, and 3.5--4 $\mu$m. As a result, the spectra at these wavelength ranges also show some sensitivity to the radiative zone \kzz\ of the deep atmosphere. This sensitivity between 2--2.2 $\mu$m, 3.2-3.5 $\mu$m, and 3.5--4 $\mu$m is shown in Figure \ref{fig:fig1pt5to1pt8} bottom panels and Figure \ref{fig:fig2to2pt5} top and bottom panels, respectively. Examining different spectral regions for a given object, that are sensitive to both \meth\ and \co, will help in constraining the radiative zone \kzz, as both gas species abundances can be used as probes. Apart from the emission spectra itself, other observables like broadband IR colors and magnitudes of brown dwarfs are also known to be sensitive to \kzz\ \citep[e.g.][]{saumon03,karilidi21}, which we explore next.

\subsection{Color Magnitude Diagrams}\label{sec:cmd}

\begin{figure*}
  \centering
  \includegraphics[width=0.9\textwidth]{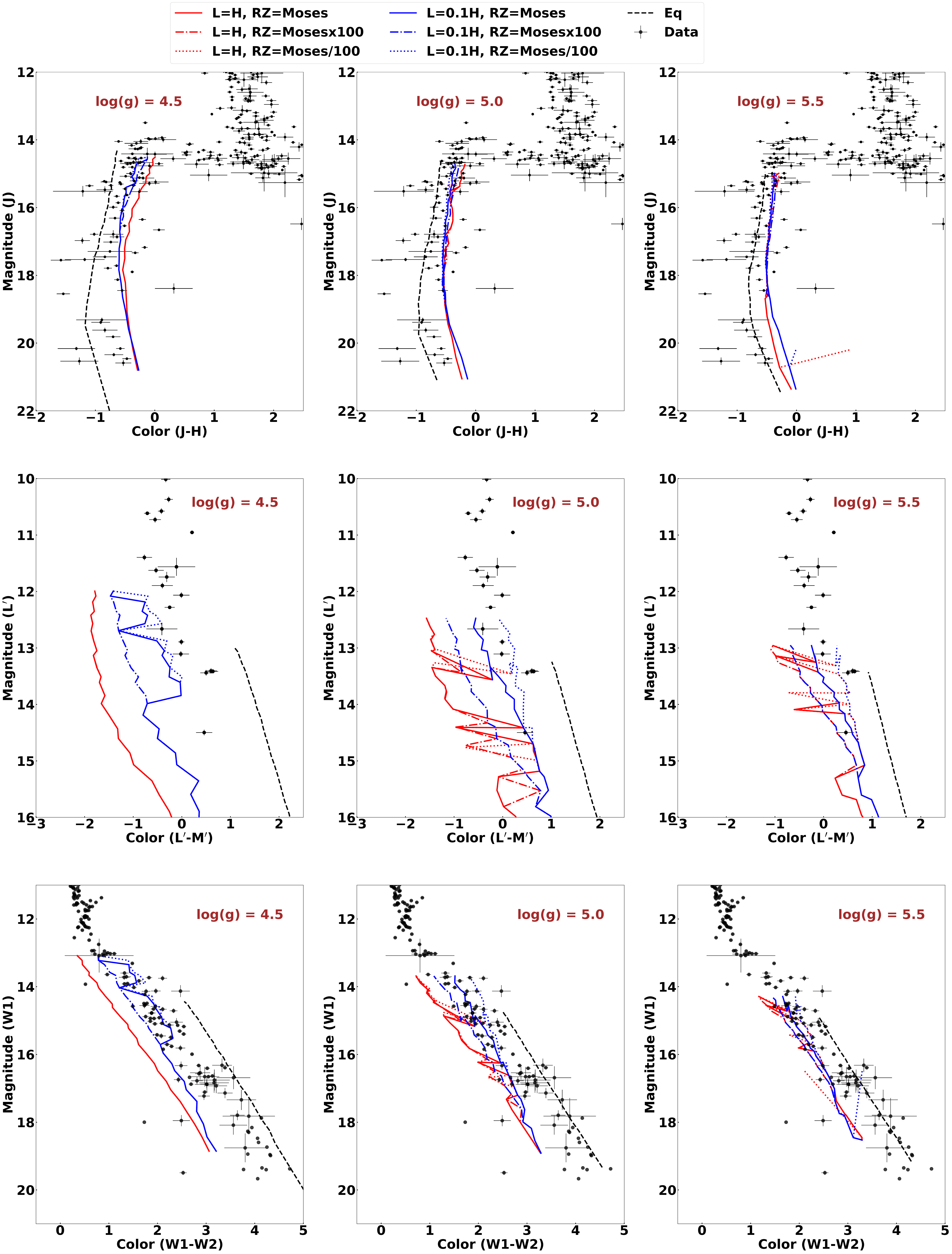}
  
  \caption{Color magnitude diagrams (CMD) of observed brown dwarfs in different filters are shown with overplotted equilibrium and disequilibrium chemistry model tracks with different strengths of convective and radiative zone vertical mixing. The top row shows the J vs. J-H where the models with convective zone mixing length of $H$ and different radiative zone {\kzz} are shown with various linetypes of red. Models with convective zone mixing length of $0.1\times{H}$ and different radiative zone {\kzz} are shown with various linetypes of blue and the equilibrium chemistry models are shown with dashed black lines. Each column shows the same data but the model tracks with different values of log(g). The second row shows the the L$^{'}$ vs L$^{'}$-M$^{'}$ whereas the third row shows the W1 vs. W1-W2.}
\label{fig:fig10}
\end{figure*}

Coupling our atmospheric model grid with brown dwarf evolutionary models from \citet{marley21} enables us to generate color magnitude diagrams (CMDs) of these brown dwarfs. These evolutionary models track the {\teff}, log(g) and bolometric luminosity as a function of age for each brown dwarf mass. The procedure is the same as in \citet{marley21}, although our models here are slightly not self-consistent, as the upper boundary condition for the \citet{marley21} cooling tracks assume equilibrium chemical abundances. 

Figure \ref{fig:fig10} shows the observed CMD of brown dwarfs in various standard infrared filter combinations. We show J vs. J-H, MKO L$^{'}$ vs. L$^{'}$-M$^{'}$, and WISE W1 vs. W1-W2 in the first, second and third rows, respectively. The mid-infrared wavelengths are most sensitive to disequilibrium chemistry effects. Model tracks for three different log(g) (4.5, 5.0 and 5.5) values are shown in the three columns for each combination of filters. The database maintained by Trent Dupuy \footnote{http://www.as.utexas.edu/~tdupuy/plx} \citep{dupuy12,dupuy13,liu16} has been used for the observational data. The red model tracks in all the panels represents models with convective mixing length set at $H$ whereas the blue tracks represents models with convective mixing length set at $0.1\times{H}$. Each line type (solid, dotted, and dash-dotted) in both of these colors represents models with the different radiative zone {\kzz} values ``Moses$\times$100", ``Moses", and ``Moses/100," respectively. The black dashed lines represents CMD tracks from equilibrium chemistry models.

In the J vs. J-H filter combination CMD, it appears that models treating disequilibrium chemistry self-consistently provide a slightly better fit to the observed photometry than the equilibrium chemistry, for models at log(g) of 5.0-5.5, representative of the field sample. The disequilibrium chemistry models with different convective mixing lengths shown by the blue and red tracks almost fall on top of each other as the J and H band fluxes are not particularly sensitive to {\co} abundance but has some sensitivity to {\meth} abundance.

The chemical equilibrium model CMD tracks follow a much redder sequence in the MKO L$^{'}$ vs. L$^{'}$-M$^{'}$ CMDs compared to the observed photometry. The disequilibrium chemistry models are comparatively much bluer. As discussed in the previous sections, enhanced \co\ abundance in disequilibrium chemistry model causes the M$'$-band and the W2-band fluxes to be fainter than equilibrium chemistry models. On the other hand, depleted \meth\ abundance in disequilibrium chemistry models causes higher fluxes in the L$'$ and W1 bands. Both of these effects causes the disequilibrium model tracks (shown with red and blue lines) to be significantly bluer than the equilibrium chemistry model tracks (shown with black line) in the L$^{'}$ vs. L$^{'}$-M$^{'}$ and  W1 vs. W1-W2 CMD. 

It is  interesting to note the non-monotonic behavior.  At higher gravity (log(g) = 5.0 and 5.5 ) some disequilibrium chemistry models show radiative zone quenching instead of convective zone quenching. As a result, these atmospheres have lower photospheric {\co} compared to disequilibrium chemistry with convective quenching, but a much higher {\co} abundance compared to the equilibrium chemistry models. Therefore, the models with radiative zone quenching are redder than disequilibrium chemistry models with convective zone quenching but bluer than equilibrium chemistry models, ``jutting out'' to the red for some models.

The second and third rows in Figure \ref{fig:fig10} also show that disequilibrium  models with $L=H$ with convective zone quenching of {\co} are much bluer than the observed sequence. However, disequilibrium models with $L=0.1\times{H}$ provide a generally better fit to the mid-IR photometry. This suggests that longer {\tmix} values in the deeper atmosphere provide a better fit to the photometry from the T and Y dwarf population, than the {\tmix} calculated from assumptions like free convection. However, this result is empirical in nature and needs further validation from 3D tracer transport models like in \citet{bordwell18}. Coupling our models with 1D chemical kinetics models will also help us in interpreting these trends better, which is something we will address in future studies. It seems clear that even lower \kzz\ values for the faintest models may be necessary, as the data are generally more consistent with equilibrium chemistry models at log (g) of 5.0-5.5  We also note that we have not included clouds within our models and they might have some impact on the model magnitudes presented here especially for the faintest end of these tracks as \water\ clouds are expected to form in the colder objects (\teff\ $\le 450$ K) \citep{morley14}.

\subsection{Comparing to Spectroscopic Observations}\label{sec:obs}

We use our grid of atmosphere models to compare with ground-based spectral observations of six late T and early Y-dwarfs. We aim to determine the atmospheric properties of brown dwarfs -- Gleise 570 D \citep{burgasser2000}, 2MASS J0415-0935 \citep{burgasser02}, WISE 0313 \citep{kirpatrick11}, UGPS 0722 \citep{cushing11}, WISE 2056 \citep{schneider15}, and WISE 1541 \citep{schneider15} with our models. Although \citet{Miles20} had the Y-dwarf WISE 0855 in their sample, we exclude this object from our analysis as it is expected to have optically thick water clouds in its atmosphere due to its cold \teff\ ($\sim$ 250 K) and our model grid only extends to a \teff\ = 400 K. As noted in \S\ref{sec:results}, this choice for the lower bound on \teff\ in our model grid was made because our models lack a self-consistent treatment of clouds with disequilibrium chemistry, currently.

For comparing our models with observational data, which includes both near-infrared and M-band spectra, we use the goodness of fit parameter -- G$_{k}$ defined as \citep{cushing08},
\begin{equation}\label{eq:gk}
    G_k = \dfrac{1}{n-m}\sum_{i=1}^{n}w_i\left(\dfrac{f_i-C_k{F^k_{i}}}{{\sigma}_i}\right)^2
\end{equation}
where $n$ is the number of observational data points, $m$ is the number of free parameters being fit, $w_i$ are the weights of fitting $i$'th data point, $f_i$ is the $i$'th observed flux, $F^k_{i}$ is the $i$'th model flux point of the $k$'th model, and $\sigma_{i}$ is the error in the $i$'th observed flux. The factor $C_k$ is a renormalization factor for the model fluxes. If the observed fluxes are flux calibrated then $C_k$ represents the factor $R^2/D^2$ but if the observed fluxes are not flux calibrated then $C_k$ represents the factor $\alpha{R}^2/D^2$ where 1/$\alpha$ represents the unknown flux normalization of the observed flux. The $C_k$ factor is given by \citep{cushing08},

\begin{equation}\label{eq:ck}
    C_k = \dfrac{\sum_{i=1}^{n}w_if_iF^k_{i}/{\sigma}_i^2}{\sum_{i=1}^{n}w_i(F^k_{i})^2/{\sigma}_i^2}
\end{equation}

Since there is no reported bias in any of the datasets we use, $w_i$ is always fixed at 1 for our analysis.

\subsubsection{Gliese 570 D and 2MASS J0415-0935}

Gliese 570 D is a T7.5 dwarf \citep{burgasser06} which has a measured spectrum in the M-band with two different instruments -- AKARI \citep{sorahana2012} and GEMINI/NIRI \citep{geballe09}. We  used the M-band spectra of Gliese 570 D reported in \citep{geballe09} because it has a higher SNR ratio. We also use the NIRC 1-2.5 $\mu$m spectra of Gliese 570 D (Gl 570D) reported in \citet{burgasser02} and {\it Spitzer} Infrared Spectrograph (IRS) 5-14 $\mu$m spectra of Gliese 570 D reported in \citet{suarez22}. We adopt the methodology described in \citet{sorahana2012} for fitting the spectrum of Gl 570D with our grid of models. Firstly, we use our grid of models to calculate $C_k$ using only the M-band GEMINI/NIRI data for Gl 570D. Using this set of calculated $C_k$, we calculate the $G_k$ parameter using our models and the M-band GEMINI/NIRI data for Gl 570D. Then we select all the models with values of the $G_k$ parameter between $min(G_k$) and $min(G_k$) +1. This step gave us 131 models which best-fits the M-band GEMINI/NIRI data for Gl 570D. The $G_k$ parameter for these 131 models are distributed between 0.9899 and 1.9550.

\begin{figure*}
  \centering
  \includegraphics[width=1\textwidth]{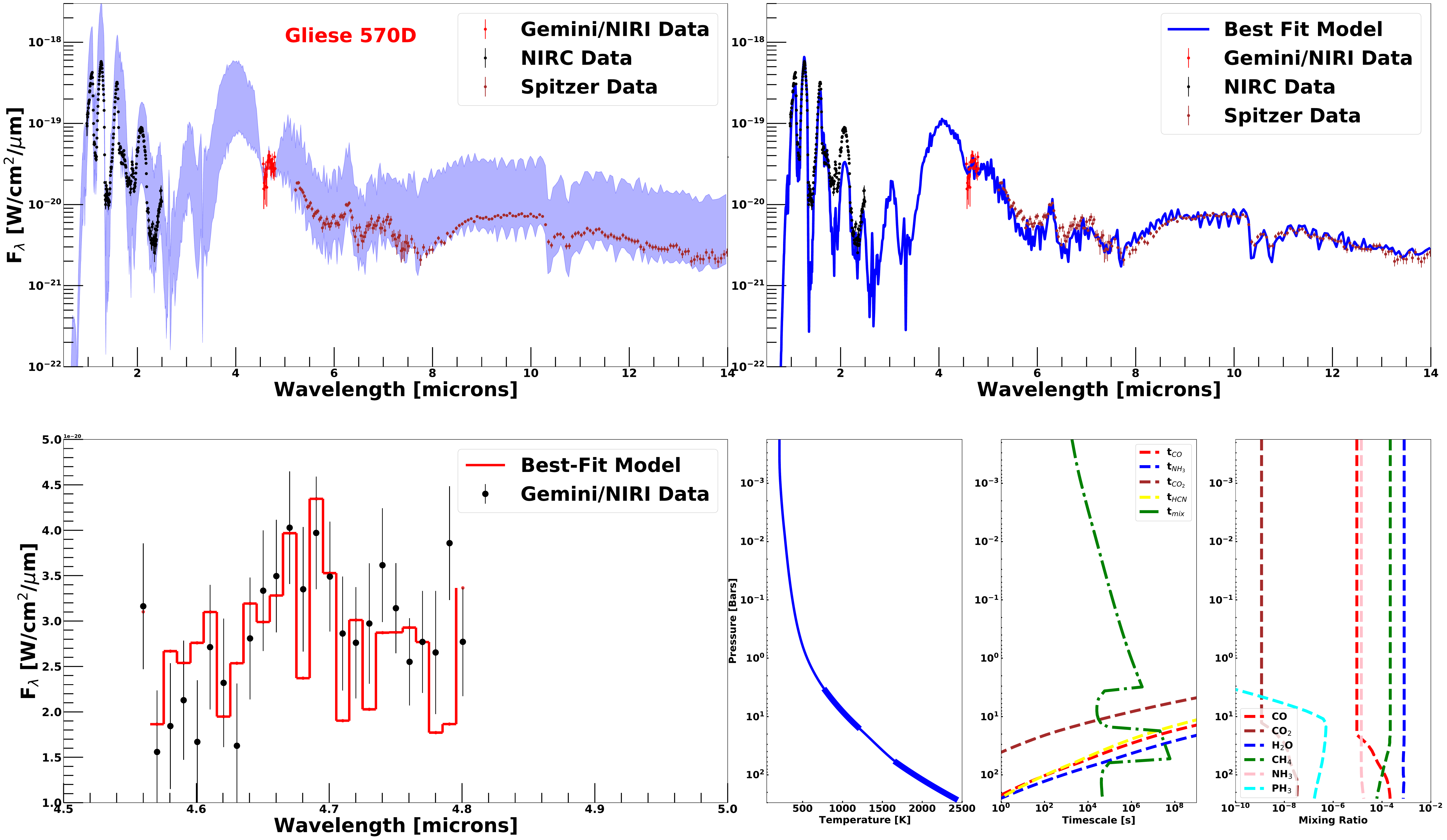}
  
  \caption{The {\bf top left panel} shows the observed M-band spectra of Gl 570D with red points, the 1-2.5 $\mu$m observations of Gl 570D with black points and the {\it Spitzer} IRS measurements with brown points.. The flux values sampled by the top 131 models with the least G$_k$ parameter after fitting the M-band spectra only are shown with the shaded blue region. The {\bf top right panel} shows the best-fit model obtained by fitting the M-band spectra, the 1-2.5 $\mu$m data, and the {\it Spitzer} 5-14 $\mu$m data for Gl 570D. The {\bf bottom left panel} shows a comparison of the best-fit model with the M-band spectra only. The {\bf bottom right panels} shows the $T(P)$ profile of the best fit model, the {\tmix} and {\tchem} of different quenched species as a function of pressure, and the rightmost panel shows the volume mixing ratios of {\co}, {\meth}, {\amon}, {\water}, {\cotwo} and PH$_3$.}
\label{fig:fig11}
\end{figure*}
\begin{figure*}
  \centering
  \includegraphics[width=1\textwidth]{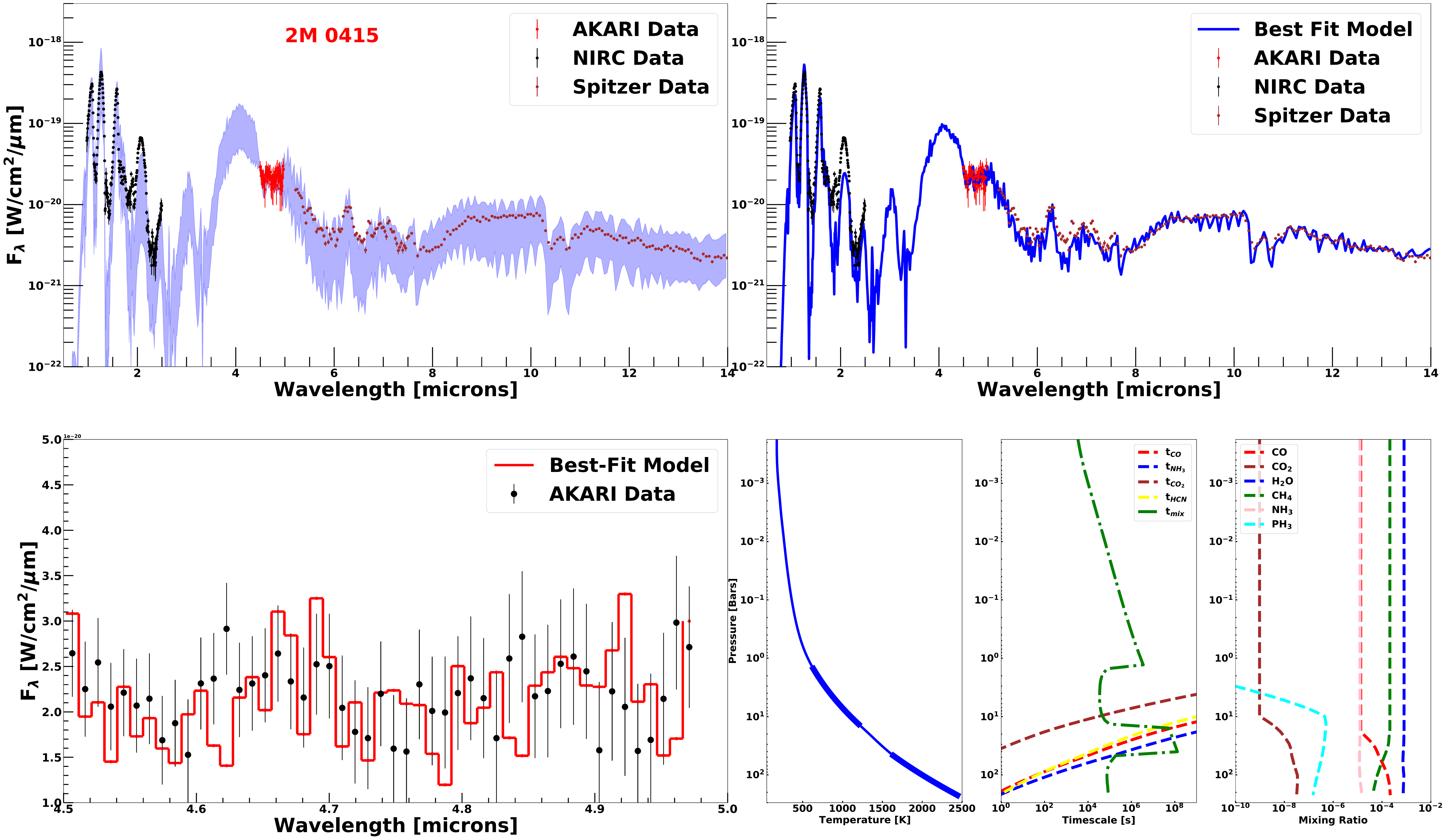}
  
  \caption{The {\bf top left panel} shows the observed M-band spectra of 2M 0415 with red points, the 1-2.5 $\mu$m observations of 2M 0415 with black points and the {\it Spitzer} IRS measurements with brown points. The flux values sampled by the top 101 models with the least G$_k$ parameter after fitting the M-band spectra only are shown with the shaded blue region. The {\bf top right panel} shows the best-fit model obtained by fitting the M-band spectra, the 1-2.5 $\mu$m data and the {\it Spitzer} 5-14 $\mu$m data for 2M 0415. The {\bf bottom left panel} shows a comparison of the best-fit model with the M-band spectra only. The {\bf bottom right panels} shows the $T(P)$ profile of the best fit model, the {\tmix} and {\tchem} of different quenched species as a function of pressure, and the rightmost panel shows the volume mixing ratios of {\co}, {\meth}, {\amon}, {\water}, {\cotwo} and PH$_3$.}
\label{fig:fig12}
\end{figure*}

The top left panel of Figure \ref{fig:fig11} shows the observed M-band spectrum of Gl 570 D \citep{geballe09} with red points, the observed NIRC spectrum of Gl 570 D \citep{burgasser02} with black points, and the {\it Spitzer} IRS spectrum \citep{suarez22} with brown points. The range of fluxes covered by the 131 best-fitting models to the M-band spectra only, are shown with the blue shaded region in the top left panel of Figure \ref{fig:fig11}. All of these models fit the M-band spectrum well but differ significantly in their fluxes in the shorter and longer wavelengths. We calculate a new set of $G_k$ parameter for these 131 models but now with the NIRC and {\it Spitzer} IRS data only. The set of $C_k$ parameters for each model, determined by fitting the M-band spectra already, are re-used for this, following the methodology of \citet{sorahana2012}. The model among these 131 models that fits this shorter and longer wavelength data as well (has the least $G_k$) is taken to be the overall best-fit model from the grid. This model spectra is shown in Figure \ref{fig:fig11} top right panel and it has a G$_k$ value of 669.06. This large difference between the G$_k$ values from M-band data and the NIRC+IRS data is mainly because of the large difference in the uncertainties in the fluxes from these two sets of observations.

The best-fit model has a {\teff} of 725 K and a log(g) of 5.0. The {\kzz} in the radiative zone for this model is Moses and the convective mixing length is $0.1\times{H}$. The M-band spectra of this model compared to the M-band data for Gl 570D is also shown in the bottom left panel. The $T(P)$ profile, mixing timescale and chemical timescales, as well as the mixing ratio profiles of various gases in this model are shown in the three bottom right panels. The $T(P)$ profile and the {\tmix} profile both show that this model has two convective zones. The {\tchem} and {\tmix} shown in the middle panel of the bottom right part of Figure \ref{fig:fig11} also show that {\co}, {\meth}, {\water}, HCN, and {\amon} are quenched in the sandwiched radiative zone for this best-fit model whereas {\cotwo} is quenched in the upper detached convective zone. The $C_k$ value of this best-fit model is 1.72$\times$10$^{-19}$. The measured parallax of Gl 570D is 171.22 mas \citep{dupuy12}. This corresponds to a distance (D) of 5.84 pc. We get a radius of 1.04 R$_{\rm J}$ for Gl 570D combining these values together. The radius of a brown dwarf with {\teff}=725 K and log(g)=5.0 can also be interpolated from evolutionary grids like the \texttt{SONORA BOBCAT} models \citep{marley21}. We find this interpolated radius to be 0.89 R$_{\rm J}$ which is somewhat smaller than our calculated radius. But as we also note in \S\ref{sec:cmd}, these evolutionary tracks were computed assuming chemical equilibrium.

We use the same technique for fitting the AKARI M-band spectrum \citep{sorahana2012} and the NIRC + {\it Spitzer} IRS spectrum \citep{burgasser02,suarez22} of 2MASS J0415-0935 (2M 0415). The top left panel of Figure \ref{fig:fig12} shows the M-band spectra of 2M 0415 with red points, the NIRC measurements are shown with black points, and the {\it Spitzer} IRS measurements are shown with brown points. The $min(G_k$) obtained by fitting the M-band spectra only is 0.8755 and the flux range covered by all the models with $G_k$ between 0.8755 and 1.8583 is shown with the blue shaded region in the top left panel. These models were then used to obtain the best-fitting model with the low and high wavelength data. The best-fit model has a $G_k$ of 514.28. This model has a {\teff} of 675 K and a log(g) of 4.75. The {\kzz} in the radiative zone for this model is Moses and the convective mixing length is $0.1\times{H}$. The measured parallax of 2M 0415 is 175.2 mas \citep{dupuy12} which corresponds to a distance of 5.707 pc. Combining this with the best-fit model $C_k$ value (1.84$\times$10$^{-19}$) provides a radius of 1.05 R$_{\rm J}$ for 2M 0415. The interpolated radius from the \texttt{SONORA BOBCAT} grid for this object is 0.98 R$_{\rm J}$. This shows much better agreement with the calculated radius of the object compared to Gl 570D. The M-band spectrum of this best-fit model compared to the M-band data for 2M 0415 is also shown in the bottom left panel. Atmospheric properties of the best-fit model are shown in the three bottom right panels. This best-fit model has a second sandwiched radiative zone and {\co}, {\meth}, {\water}, and HCN are quenched in this sandwiched radiative zone whereas {\cotwo} is quenched in the upper detached convective zone.

\subsubsection{WISE 0313, UGPS 0722, WISE 2056 and WISE 1541}

\begin{figure*}
  \centering
  \includegraphics[width=1\textwidth]{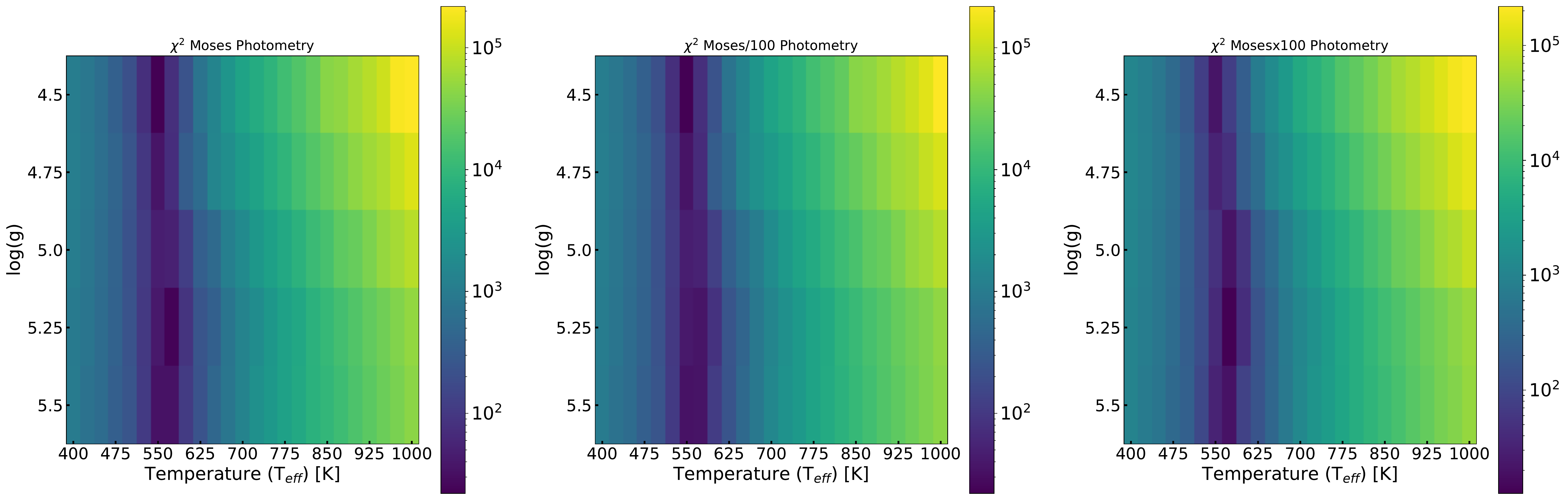}
  
  \caption{The {\bf left panel} shows a heat map of the $\chi^2$ value calculated using model photometry and observed photometry \citep{Miles20} in various infrared bands as a function of model {\teff} and log(g) for WISE 0313. The left panel indicates models where radiative {\kzz} is fixed at Moses and the convective mixing length is ${H}$. The {\bf middle and right panels} also show the $\chi^2$ heat maps using models with radiative {\kzz} of Moses/100 and Moses$\times$100, respectively. The best-fit {\teff} is near 550 K.}
\label{fig:fig13}
\end{figure*}

\citet{Miles20} has obtained M-band observations of the late T and early Y-dwarfs WISE 0313, UGPS 0722, WISE 2056, and WISE 1541. However, these M-band spectra are not flux calibrated. Using photometry collected from various studies and the \texttt{SONORA} evolutionary tracks, \citet{Miles20} found ranges of {\teff} and log(g) values for these objects. We use our models along with these photometry datasets (see Table 2 in \citet{Miles20} for photometry obtained from \citet{skrutskie06,kirpatrick11,lucas10,leggett13,schneider15,geballe01,knapp04,leggett15,golimowski04,leggett17,patten06,kirkpatrick12,cutri14,luhman16,schneider16,wright14,esplin16}) to find a similar range of {\teff} values for each of these sources. Figure \ref{fig:fig13} shows the map of ${\chi}^2$ values in a {\teff} vs. log(g) parameter space obtained by comparing our models with observed photometry of WISE 0313. It is clear that the ${\chi}^2$ values are sensitive to the {\teff} of models but there is not much sensitivity to the log(g) of the models. This method helps us in identifying a range ($\delta$T = 100 K) of {\teff} values for each object. Then using models within this {\teff} range we try to determine the best-fit atmospheric model by comparing our model spectra shapes with the shape of the M-band spectra of these four objects. We use the same metric $G_k$ and $C_k$ defined in Equations \ref{eq:gk} and \ref{eq:ck} for finding the best-fit models. However, as the observed spectra lacks flux calibration, the $C_k$ now includes the $\alpha$ parameter as well where 1/$\alpha$ represents the unknown flux calibration of the observed spectrum. For each object, we use the models within the determined {\teff} ranges and allow all the other parameters such as log(g), radiative zone {\kzz} and mixing length $L$ to vary while calculating the $G_k$ and $C_k$ values of the models using the M-band spectrum. Then the model with the least $G_k$ value is chosen as the best-fit model for the object.

\begin{figure*}
  \centering
  \includegraphics[width=1\textwidth]{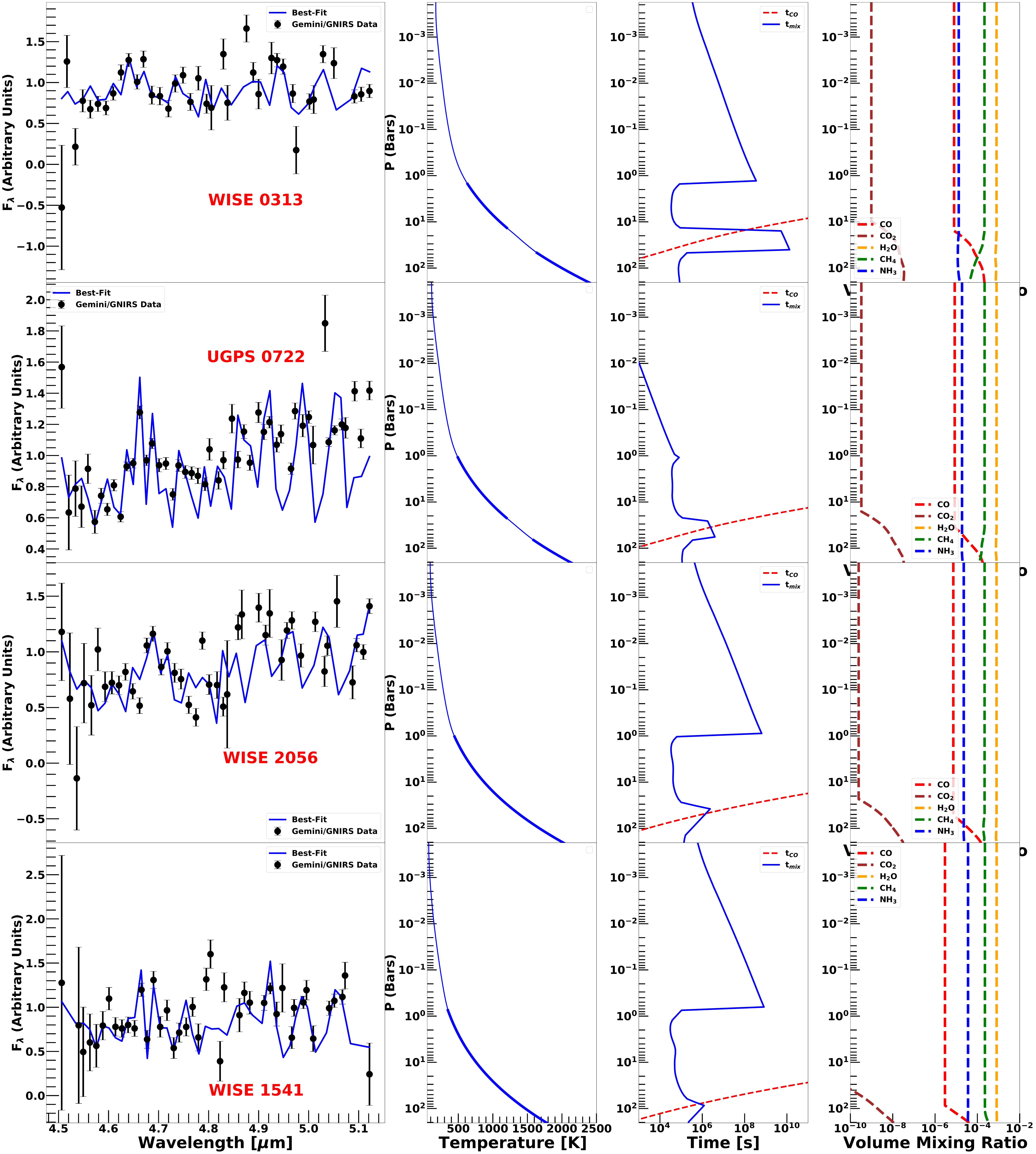}

  \caption{The best-fit atmospheric models and their comparison with data are shown in this figure. {\bf Each row} corresponds to a different object. The {\bf left most column} shows the best-fit M-band spectra along with the observed data from \citet{Miles20}. The {\bf second left column} shows the best-fit $T(P)$ profile for these objects from the grid of models presented in this paper. The {\bf third column shows} the best-fit mixing timescale in blue along with the chemical timescale for {\co} in red as a function of pressure. The volume mixing ratios of various atmospheric gases are shown in the {\bf fourth column} for each object.}
\label{fig:fig14}
\end{figure*}

  
  

Figure \ref{fig:fig14} shows the results of this fitting exercise for all the four objects. The four rows in Figure \ref{fig:fig14} shows the results for WISE 0313, UGPS 0722, WISE 2056, and WISE 1541, respectively. The left columns show the best-fit spectra compared with the observational data for each object. The second column from left shows the best-fit $T(P)$ profile for each object. The third column shows the best-fit mixing timescale and chemical timescale for {\co} as a function of pressure for these objects and the fourth columns depicts the best-fit volume mixing ratios for each object.

We find that the best-fit model describing the WISE 0313 and UGPS 0722 data have radiative zone quenching of {\co}/{\meth} whereas the best-fit models describing the WISE 2056 and WISE 1541 data has {\co} quenching in the convective zone. Table \ref{tab:tab1} shows the values of all the parameters relevant to fitting the spectral data for all of these objects. We note that the best-fit log(g) values obtained in this work are not from fitting any spectral shape/feature directly. As can be seen in Figure \ref{fig:fig4}, log(g) controls the presence/absence of a sandwiched radiative zone which in turn dictates whether gases can be quenched in regions with low or high {\kzz}. These effects of log(g) on atmospheric chemistry lead to preference of certain log(g) values over others while fitting the data.
\begin{table*}\label{tab:tab1}
\begin{center}
\begin{tabular}{||m{1.5cm} m{1.5cm} m{1.5cm} m{1.5cm} m{1.5cm} m{1.5cm} m{1.5cm} m{3cm} m{2cm} ||} 
 \hline
 Source & {\teff} Range[K] & Best-Fit {\teff}[K] & Best-Fit log(g) & Radiative {\kzz} & Convective L & M-band G$_k$ & J--H--K+\emph{Spitzer}IRS G$_k$ & C$_k$ \\ [0.5ex] 
 \hline\hline
 Gl 570D & 700--800 & 725 & 5.0 & Moses & $0.1\times{H}$ & 1.414 & 669.06 & 1.72$\times$10$^{-19}$ \\ 
 \hline
 2M 0415 & 600--700 & 675 & 4.75 & Moses & $0.1\times{H}$ & 0.925 & 514.28 & 1.84$\times$10$^{-19}$ \\
 \hline
 WISE 0313 & 575--675 & 675 & 4.75 & Moses/100 & $0.1\times{H}$ & 3.849 & -- & 6.477\\
 \hline
 UGPS 0722 & 500--600 & 575 & 4.75 & Moses$\times$100 & $0.1\times{H}$ & 23.884  & -- & 11.636 \\
 \hline
 WISE 2056 & 450--550 & 525 & 4.75 & -- & $0.1\times{H}$ & 6.651 & -- & 14.575 \\
 \hline

 WISE 1541 & 400--450 & 450 & 4.75 & -- & $0.1\times{H}$ & 5.243 & -- & 20.275\\

 \hline
\end{tabular}

\end{center}
\caption{Results from fitting observed spectra of brown dwarfs with our model spectra. The radiative zone {\kzz} for models which do not quench at radiative zones are marked with a '--' as they couldn't be constrained.}
\end{table*}

We choose our best-fit model to be the model with the least $G_k$ value but we also find that there are quite a few models which have $G_k$ values only differ from the minimum only by a small amount. This shows the presence of degeneracy in our models with the current precision of available data for these objects. This degeneracy will be broken with much higher precision flux calibrated M-band and shorter wavelength spectra expected from JWST which we discuss in \S\ref{sec:jwst}.

\subsubsection{Trends in Quench Timescales}
\begin{figure*}
  \centering
  \includegraphics[width=1\textwidth]{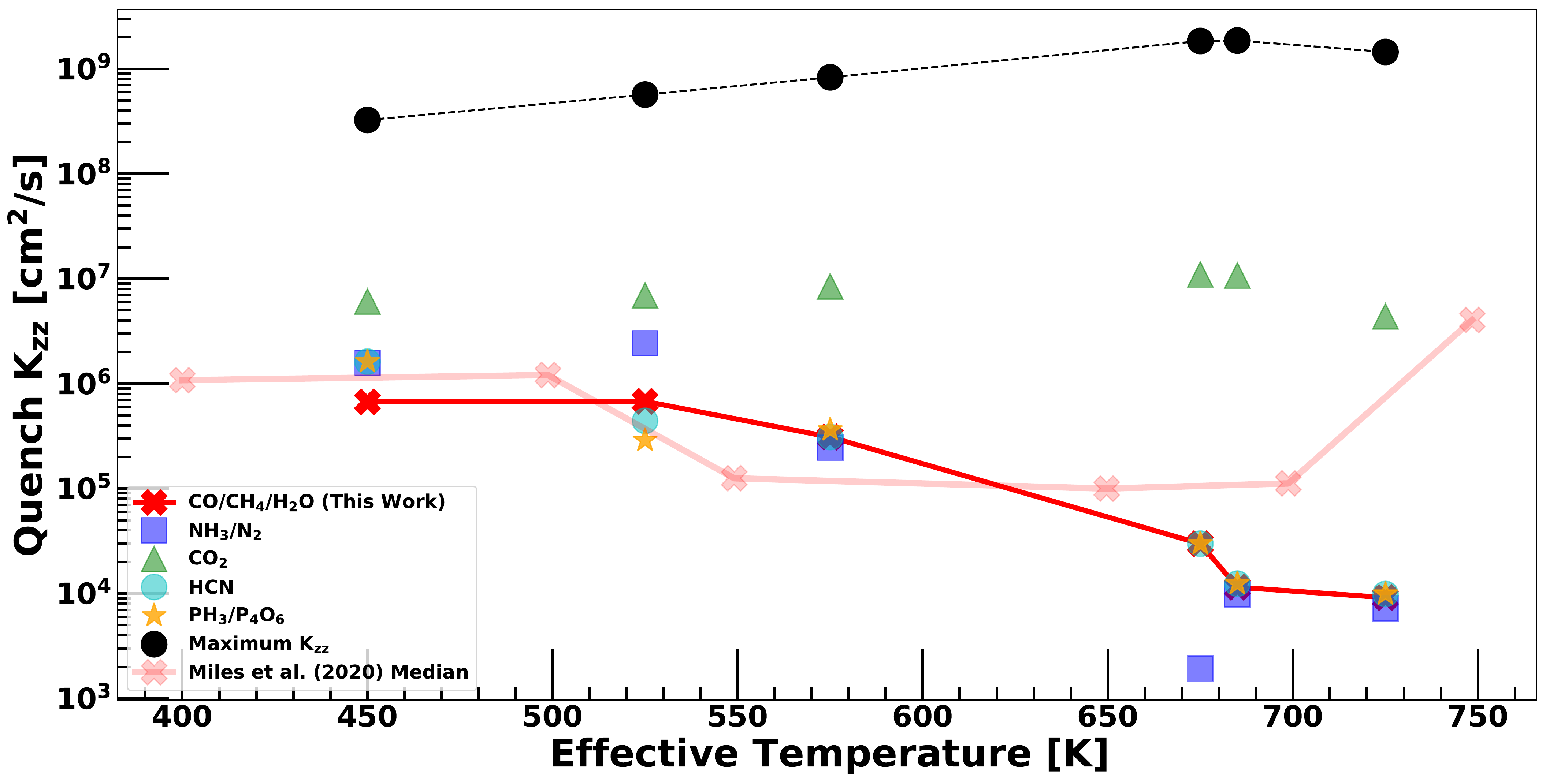}

  \caption{The best fit quench {\kzz} of {\co} for each object is shown as a function of their best-fit {\teff} with the red crosses. The median best-fit quench {\kzz} vs. {\teff} inferred by \citet{Miles20} with lower gravity models are shown with the faded red crosses. The inferred quench {\kzz} for {\amon}, {\cotwo}, HCN, and \phos\ from the best-fit models for each object are also shown as a function of their best-fit {\teff}. The black points show the theoretically maximum {\kzz} allowed from mixing length theory for these best-fit models.}
\label{fig:fig16}
\end{figure*}

We present the trend of the best-fit {\kzz} at {\pq} for \co\ in these sources in Figure \ref{fig:fig16}. As we mainly fit the M-band spectra of these objects in our analysis, we are only sensitive to \co\ and \water\ abundances and therefore we can only fit the quench {\kzz} of {\co}. Our best-fit quench \kzz\ for \co\ as a function of the best-fit \teff\ for each object are shown with red crosses in Figure \ref{fig:fig16}. The best-fit quench \kzz\ determined by \citet{Miles20} are shown with faded red crosses as a function of the best-fit \teff\ obtained by \citet{Miles20} for each object. Although the overall trend of much lower \co\ quench \kzz\ than the theoretical maximum in objects with \teff\ between 600--800 K, found by \citet{Miles20}, still stands in this analysis, there are some differences in the best-fit \kzz\ and \teff\ values between the two studies. These differences are due to multiple factors which we discuss briefly here. 

Our work uses RCE models that include disequilibrium chemistry self-consistently to fit the data while equilibrium chemistry RCE models were post-processed by \citet{Miles20} to include quenching of gases to fit the data. Figure \ref{fig:fig2} top left panel shows that including disequilibrium chemistry self-consistently can lead to large changes in the $T(P)$ profile of brown dwarfs compared to equilibrium chemistry models. This can be the reason behind the $\sim$ 50 K differences in the \teff\ values for these objects between the two studies. For 5 of the 6 analyzed objects in this work, our assessment of \co\ quench \kzz\ are somewhat smaller than previously found. As some of our best-fit models have higher \teff\ than the best-fit models of \citet{Miles20}, \co\ is more abundant in the deeper atmosphere of our best-fit models than those of \citet{Miles20}. Figure \ref{fig:fig2} shows that \co\ abundance falls very rapidly with decreasing pressure in the deep atmosphere (which is under chemical equilibrium) in these colder brown dwarfs. Therefore colder best-fit models in \citet{Miles20} need to quench \co\ deeper in the atmosphere than our hotter best-fit models. So, the best-fit models presented in \citet{Miles20} need higher \co\ quench \kzz\ compared to our best-fit models to produce the same \co\ abundance in the upper well mixed atmosphere of these brown dwarfs. Additionally, the best-fit \kzz\ presented in \citet{Miles20} has been shown to be dependant on the gravity of the model used for fitting the data. In this work, we  allowed gravity to be a free parameter while fitting the data and  compared our best-fit \kzz\ in Figure \ref{fig:fig16} with the median \kzz\ obtained by \citet{Miles20} by trying out various values of log(g). 

We can use our best-fit models with \co\ quenching to infer the quench \kzz\ of other quenched gases like {\cotwo}, {\amon} or HCN in these best-fit models. The inferred quench \kzz\ values for the other gases in these objects are also shown in Figure \ref{fig:fig16}. As each gas has a different {\tchem}, they also are quenched at different depths of the brown dwarf atmosphere as shown in Figure \ref{fig:fig5} and \ref{fig:fig7} for \co\ and {\cotwo}. This means that measuring the abundance of each gas can help us in measuring the height dependant {\kzz} profile in these brown dwarf atmospheres. For example, in Figure \ref{fig:fig16}, the inferred quench \kzz\ for \cotwo\ are very different than the inferred quench \kzz\ of \co\ (or {\amon}) as \cotwo\ is quenched at different pressures (having different mixing strengths) compared to {\co}. But inferring the height dependant {\kzz} profile of a single brown dwarf by using different gases as probes of \kzz\ at different pressures is only possible with self-consistent disequilibrium chemistry models with height dependant \kzz\ profiles such as the model grid presented in this work.

The theoretical maximum limit on convective zone {\kzz} for each best-fit model is also shown with black points in Figure \ref{fig:fig16}. This maximum limit on convective {\kzz} was calculated using Equation \ref{eq:kzz_con} with the convective heat flux fixed at $\sigma${\teff}$^4$ and the mixing length $L$ fixed at the scale height of the atmosphere. It is clear from Figure \ref{fig:fig16} that all the gases in all these objects quench at lower {\kzz} than this theoretical limit. But even though all the objects analyzed in this work have much smaller \co\ quench {\kzz} than the maximum convective limit shown with the black line, this does not mean that all these ``quenchings'' occurred in the radiative zones in our models. For Gl 570D, 2M 0415, W0313, and UGPS 0722, {\co} is quenched in the radiative zones in our best-fit models but for W 2056 and W 1541 {\co} along with all the other molecules are quenched in the convective zone with a much smaller mixing length than the scale height of the atmosphere as was also suggested by the CMD analysis detailed in \S\ref{sec:cmd}. Finally, we note that our best-fit models, especially for objects which lack flux-calibrated spectra, are often degenerate with the present data, and better data from observatories like JWST will be instrumental in giving us better constraints on the {\kzz} of these objects, which we discuss next.

\subsection{Constraining Radiative Zone {\kzz} with JWST}\label{sec:jwst}

\begin{figure*}
  \centering
  \includegraphics[width=1\textwidth]{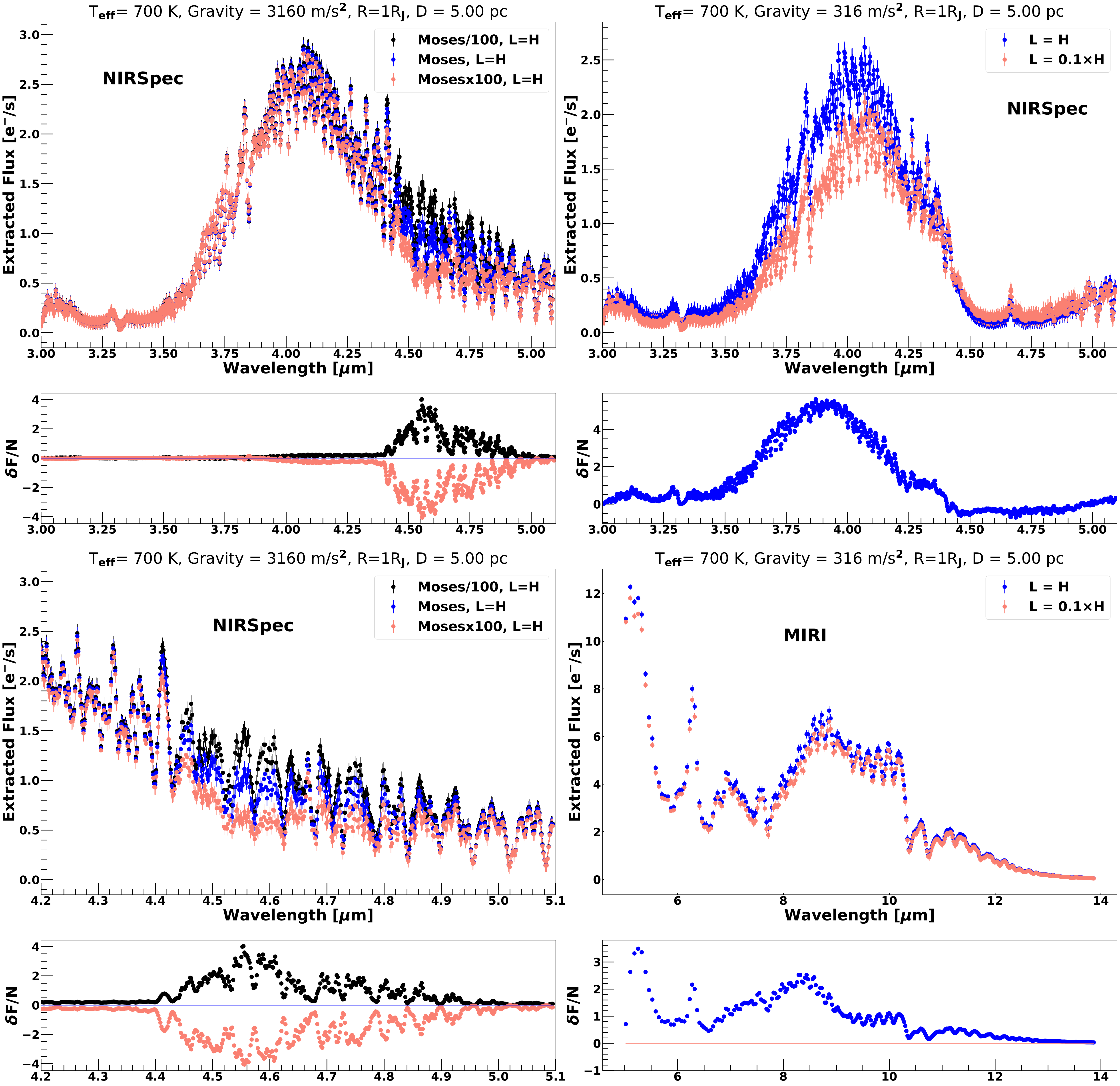}

  \caption{Simulated data from JWST instruments are  shown for one of our model brown dwarfs with {\teff} of 700 K. The three colors in the simulated data correspond to models with different radiative zone {\kzz}. The {\bf top left panel} shows the simulated NIRSpec spectra of a log (g)=5.5 object with an exposure time of 10 minutes while the {\bf bottom left panel} shows a zoomed version of the top left panel to emphasize how JWST NIRSpec M-band observations can clearly distinguish between different {\kzz} in the radiative zones for these T-dwarfs. The {\bf top right panel} shows the NIRSpec simulated data for a brown dwarf with {\teff} of 700 K and log(g) of 4.5. The two colors represent different {\kzz} in the convective zone but the same {\kzz} in the radiative zone. The {\bf bottom right panel} shows MIRI LRS simulated data for the same models shown in the top right panels. To compare the differences between the simulated data from different models to the typical JWST noise, the ratio of the flux differences ($\delta$F) between various models and the simulated JWST noise ($N$) is shown below each of these panels. This shows that both convective and radiative zone {\kzz} can be constrained with JWST.}
\label{fig:fig17}
\end{figure*}

An important takeaway from \citet{Miles20} and our work is that detailed analysis of atmospheric abundances can constrain \kzz\ in brown dwarf atmospheres.  There is strong evidence for mixing orders of magnitude weaker than expectations from free convection with $L=H$.  The launch of JWST begs the question of whether JWST data will be able to give us better constraints on the vigor of vertical mixing in these atmospheres. We can provide a test by simulating the JWST spectra using the JWST Exposure Time Calculator \citep{pontoppidan16} with some of the emission spectra from our model grid. As an illustrative example, we use the spectrum of a 700 K object.  In the left panels of Figure \ref{fig:fig17} we examine three models at log(g) of 5.5, where the radiative zone {\kzz} is ``Moses$\times$100", ``Moses", and ``Moses/100" and $L=H$. We assume that the source has a radius of 1R$_{\rm J}$ and is at a distance of 5 pc. The temperature and distance chosen for this exercise are somewhat representative of objects like Gl 570D and 2M 0415.

The NIRSpec panels (which is 3 of the 4 figures) assume fixed slit mode with the G395M grating and the F290LP filter for these simulations. The results are impressive, with only a total of 10.39 minutes exposure time, simulated with 10 groups per integration, 2 integration per exposure, and 5 exposures. It is clear from the top left panel that the radiative zone {\kzz} can be constrained with high precision with about a total $\sim$ 10 minutes exposure time per object in this JWST observing mode. This time includes all the 5 exposures used for this simulation. In order to emphasize the difference arising in the synthetic counts from the three models in the 4.5-5 $\mu$m range, the bottom left panel shows a zoomed in view of the 4.3-5 $\mu$m wavelength range, focusing on \co\ absorption, where the main sensitivity of the data to radiative zone {\kzz} lies.  The top right panel shows a lower-gravity model (log (g)=4.5), where a lower pressure photosphere favors CO at the expense of {\meth}.  Here the blue model is simulated from a model where the convective mixing length $L=H$ whereas the salmon colored data is simulated from a model where the convective mixing length is $L=0.1\times{H}$. The radiative zone {\kzz} of both the cases in this panel are ``Moses$\times$100". Both of these models have the quenching of all relevant gases occurring in the convective zone. The large differences in the synthetic counts between the two models in the 3.5--4.25 $\mu$m range shows that JWST data can also be used to probe differences in the convective zone {\kzz} with exposure times of $\sim$ 10 minutes.

Radiative and convective zone {\kzz} can be probed with other wavelengths as well. The bottom right panel in Figure \ref{fig:fig17} shows the simulated spectrum of the two cases shown in the top right panel but when observed by the MIRI instrument. The MIRI LRS mode was used for this data simulation along with a 60 minutes exposure time. Differences between the two model spectra arise between 5--9 $\mu$m. A comparison with Figure \ref{fig:fig1} shows that these differences mainly arise from different {\meth}, {\water} and {\amon} abundances due to differences in the convective zone {\kzz}.

\section{Discussion}\label{sec:disc}

\subsection{{\phos} Quenching}

\begin{figure}
  \centering
  \includegraphics[width=0.6\textwidth]{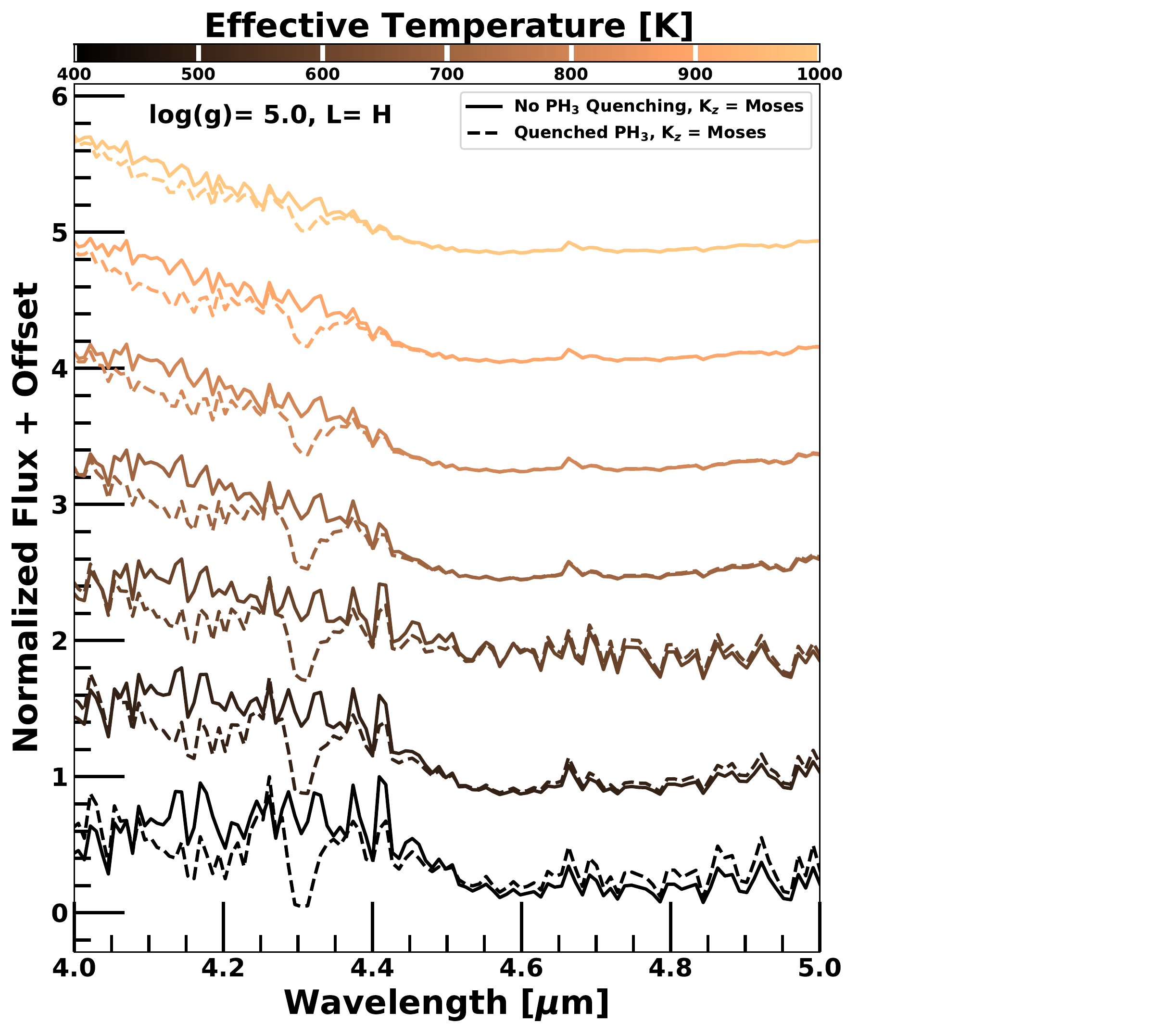}

  \caption{Effect of {\phos} quenching on the spectrum has been shown by comparing them with disequilibrium chemistry model spectra where {\phos} is not quenched. Models with different \teff\ values ranging from 400 K to 1000 K, with a step of 100 K, are shwon from bottom to top. All of these models have log(g)=5.}
\label{fig:fig18}
\end{figure}

\begin{figure}
  \centering
  \includegraphics[width=0.45\textwidth]{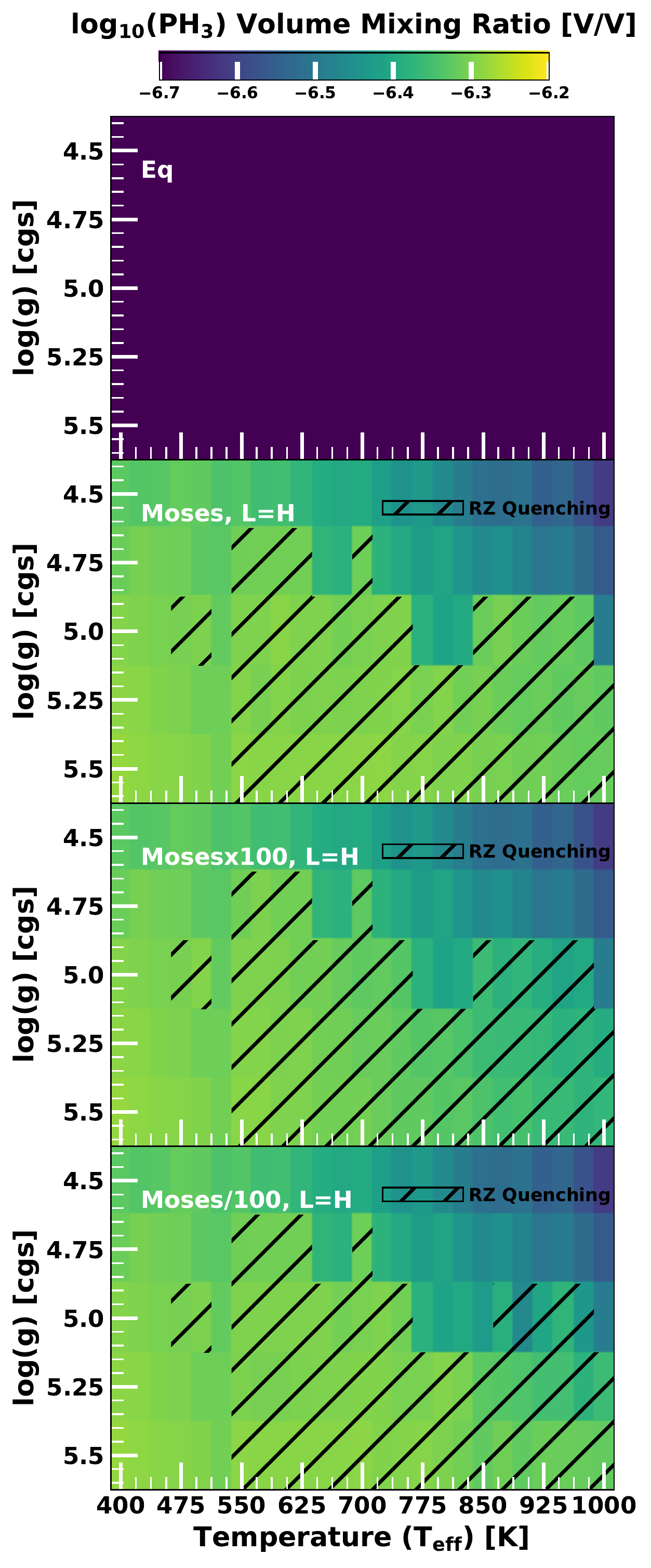}

  \caption{{\bf Top panel} shows the equilibrium chemistry abundance of PH$_3$ at pressures where $T(P)$= \teff\ in our \teff\ vs. gravity grid of chemical equilibrium models. The {\bf second, third and last panels} depict PH$_3$ abundance within the \teff\ vs. gravity parameter space in our disequilibrium chemistry models. The second, third and last panels correspond models with radiative zone \kzz\ values of ``Moses", ``Moses$\times$100", and ``Moses/100", respectively while the convective zone mixing length $L$ was fixed at the scale height $H$. Models where PH$_3$ was quenched in a radiative zone are marked with black hatches. PH$_3$ was quenched at a convective zone in the rest of the models.}
\label{fig:fig19}
\end{figure}

Under chemical equilibrium conditions the major phosphorus-bearing molecule PH$_3$ is expected to be removed by oxidation at low temperatures in the visible atmospheres of giant planets and colder brown dwarfs. A number of oxidation products have been considered \citep[see][]{visscher2020}, including P$_4$O$_6$ \citep[e.g.,][]{fegley1994,visscher06}, P$_4$O$_{10}$ \citep{borunov1995}, H$_3$PO$_4$ \citep{wang2016}, or condensation of NH$_4$H$_2$PO$_4$ \citep{fegley1994,visscher06,morley2018}. Following the work of \citet{visscher05,visscher06,Miles20}, as a representative example of PH$_3$ oxidation we may consider the conversion of PH$_3$ into P$_4$O$_6$ via the net thermochemical reaction,
\begin{equation}
    {\rm P_4O_6 + 12H_2 \rightleftarrows 4PH_3 + 6H_2O} 
\end{equation}
However, the {\tchem} of {\phos}$\rightarrow$P$_4$O$_6$ is such that it can be quenched in the deeper atmosphere, leading to a large enhancement of {\phos} in the photosphere. This effect makes {\phos} the major phosphorus bearing gas in atmospheres of Jupiter and Saturn \citep{lodders04,visscher05}. {\phos} is expected to be found in colder brown dwarfs as well \citep{visscher06,morley2018}.

Due to expected small affect of P-bearing molecules on brown dwarf temperature structures, we did not include {\phos} as a molecule to be mixed on--the--fly for converging to a final atmospheric state.  However, in order to test the effect of {\phos} on the emergent thermal emission spectra, we post-process our converged disequilibrium chemistry model with quenched {\phos} abundance. The {\tchem} for {\phos}$\rightarrow$P$_4$O$_6$ is given by \citet{visscher05},

\begin{equation}\label{eq:phos}
    t_{\rm chem} = \dfrac{1.9\times10^{12}}{[\rm OH]}{\rm exp}\left(\dfrac{6013.6}{T}\right) {\rm s}
\end{equation}
where {[OH]} represents the number density of OH molecules in units of cm$^{-3}$. With our converged disequilibrium chemistry modules, we calculate the mixing ratio of OH molecules using \citet{visscher05},

\begin{equation}\label{eq:oh}
    X_{\rm OH} = \dfrac{KX_{\rm H_2O}\sqrt{X_{\rm H_2}}}{\sqrt{P}}
\end{equation}
where X$_{\rm OH}$, X$_{\rm H_2O}$, and X$_{\rm H_2}$ represent the volume mixing ratios of each of these components and $P$ represents the pressure in bars. The constant $K$ is temperature dependant and based on fits to thermodynamic data, and is given by,

\begin{equation}\label{eq:K}
    K = 3.672 - \dfrac{14791}{T}
\end{equation}
where $T$ is the temperature in K. Using Equations \ref{eq:phos}, \ref{eq:oh}, and \ref{eq:K} we determine the {\tchem} of {\phos} for all our converged models. Then using our converged {\tmix} profiles we determine {\pq} for {\phos} and quench it creating another set of our models which has quenched {\phos} abundances. Using these modified chemical compositions, we create another set of thermal emission spectra models with quenched {\phos} abundances. Figure \ref{fig:fig18} compares the differences in spectra caused by the quenching of {\phos} with disequilibrium models where {\phos} is not quenched. This quenching of {\phos} increases the {\phos} volume mixing ratios dramatically, from $\le$ 10$^{-30}$ under chemical equilibrium, to $\sim$ 10$^{-6}$ to 10$^{-7}$ due to quenching. The main impact of this increase appears between 4-4.5 $\mu$m in the thermal emission spectrum as shown in Figure \ref{fig:fig18}. This also suggests that {\phos} may become an important enough atmospheric opacity source to have a small effect on their $T(P)$ profiles as well, which we will aim to explore in future work.

Figure \ref{fig:fig19} shows the parameter space of models where {\phos} quenches at the radiative zone and convective zones for different levels of radiative zone {\kzz} when the convective mixing length is $H$. This also shows that depending on the properties of the object in question, {\phos} can also be used as a probe for determining radiative or convective zone {\kzz}. The region around 4.0-4.4 $\mu$m is difficult to probe from the ground \citep{Miles20}, but it is clear from Figure \ref{fig:fig18} that these {\phos} should be readily detectable with \emph{JWST}. Interestingly, \citet{morley2018} didn't detect {\phos} between 4.0-4.4 $\mu$m in WISE 0855 even though it is expected to be present in it's atmosphere theoretically.  Determining the strength of these features, or their absence, will have important implications in understanding {\kzz} and phosphorus chemistry.

\subsection{Clouds}
The models presented in this work are cloud-free. However, {\water} condensation in the upper atmospheres of objects can begin at {\teff} $\le$ 450 K \citep{morley14} and these clouds can start becoming an important opacity source colder than $\sim$ 400 K \citep{morley14}. Once water clouds become optically thick, they can change the overall energy balance of the atmosphere impacting the $T(P)$ profile. The presence of disequilibrium chemistry causes a cooling down of the upper atmosphere, compared to equilibrium chemistry, but the opacity of clouds has the opposite effect of heating up the $T(P)$ profile. 

Importantly, the cloud particle size and vertical extent are also quite sensitive to the mixing parameter {\kzz} as mixing helps to keep cloud particles aloft against their gravitational settling \citep{ackerman01}. Cloud particle sizes are an extremely important parameter in dictating their scattering properties \citep{mukherjee21,ackerman01} and hence {\kzz} has a large effect on the cloud opacities in these atmospheres. Therefore, it is clear that information gleaned on {\kzz} from disequilibrium chemistry studies will be important in improving cloud models.  Future studies should aim for a self-consistent treatment of disequilibrium chemistry and cloud formation with a self-consistent treatment of {\kzz} within a given model atmosphere.  Recently, \citet{mang22} showed that water clouds have a large impact on the M-band spectra of colder Y-dwarfs, with {\teff} $\le$ 400 K, just below the lower $T$ edge of our study. Somewhat different than this work, \citet{mang22} calculated the radiative zone {\kzz}, needed for cloud modeling, by setting the mixing length to one--tenth of the scale height in the radiative zones. This approach also leads to smaller {\kzz} in the radiative zones compared to the convective zones. When we extend our disequilibrium chemistry model to colder objects, inclusion of clouds in this framework is an important milestone.

\section{Summary and Conclusion}\label{sec:sum}

In this work, we have used a newly developed Python--based 1D radiative-convective equilibrium model with the capability to self-consistently capture disequilibrium chemistry in brown dwarf and exoplanet atmospheres. Using this model we have created a self-consistent grid of brown dwarf models between {\teff} of 400-1000 K with an increment of 25 K and log(g) of 4.5-5.5 with an increment of 0.25. Instead of using a constant {\kzz} prescription throughout the entire atmosphere as done in earlier work \citep[e.g.][]{karilidi21,Philips20} or self-consistent {\kzz} in the convective zones and constant {\kzz} in the radiative zones \citep[e.g.][]{Hubeny07}, we use a self-consistent $T(P)$ structure dependant {\kzz} profile for the entire atmosphere for our grid. In the convective zone, we use mixing length theory with two different mixing lengths whereas in the radiative zones we use various multiples of a $T$ and $P$-dependent {\kzz} prescription. The idea behind these models was to assess whether disequilibrium chemical abundance can be used as a probe for temperature structure and mixing processes below the visible atmosphere in brown dwarfs. With this grid in hand, we conclude on the following points.  

\begin{enumerate}
    \item As found by others, a self-consistent treatment of disequilibrium chemistry causes cooling of $T(P)$ profiles, especially in the upper atmospheres of late T and early Y-dwarfs.
    
    \item Post-processing equilibrium chemistry models by quenching various molecules instead of using self-consistent models with disequilibrium chemistry should be avoided because this can lead to significant errors in the calculated abundances of molecules like {\co}, {\meth} and {\amon}. This can also lead to incorrect $T(P)$ profiles. Both of these results are especially relevant in the JWST era with significantly higher S/N mid-infrared data.
    
    \item Self-consistent models including disequilibrium chemistry also show a large change in the location, depths, and number of radiative and convective zones in brown dwarfs compared to self-consistent models with equilibrium chemistry.
    
    \item Depending on the strengths of radiative zone {\kzz} and convective mixing length, gases like {\co}, {\meth}, {\amon}, {\cotwo} and {\phos} can be quenched in the radiative zones of brown dwarfs in a large fraction of the {\teff} and log(g) parameter space. 
    
    \item Since different molecules have different \tchem\ values, which leads them to quench at different pressure levels.  This means that some molecules may quench in radiative zones, while other may quench in convective zones, in the same atmosphere, such that different molecules may be quenched by very different \kzz\ values.
    
    \item The M-band thermal spectra of objects where radiative zone quenching occurs, is very sensitive to the {\kzz} in the radiative zones.  This can lead to non-monotonic behavior in the \co\ abundance and spectra as a function of \teff.  This is especially exciting because it provides an avenue to observationally constrain the highly uncertain radiative zone {\kzz} in these objects.
    
    \item The mid-infrared photometric bands -- MKO L$^{'}$ and MKO M, and WISE 1 and 2, are quite sensitive to the {\kzz} in the radiative zone. We also find that our models provide a better fit to the infrared photometry of these objects than chemical equilibrium models.
    
    \item We use our models to fit the M-band spectra of 6 brown dwarfs -- Gleise 570 D, 2MASS J0415-0935, WISE 0313,  UGPS  0722, WISE  2056,  and  WISE 1541. We determine the best-fit atmospheric state for these objects including the quench level {\kzz} of quenched gases in these objects. Similar to \citet{Miles20}, we also find that these objects have {\kzz} much smaller than the theoretical convective mixing limit. We also see that the best fit quench {\kzz} in these objects are very low between {\teff} of 675-800 K while it shows an increase with lowering {\teff} for objects with {\teff} below 575 K.
    
    \item By performing some signal--to--noise calculations for JWST, we conclude that short exposure time observations can provide us with excellent measurements of the radiative and convective zone {\kzz} in brown dwarfs.
    
    \item We also expect JWST to find clearly detectable disequilibrium {\phos} abundances in the atmospheres of these objects.
\end{enumerate}

 The non-monotonic behaviour of atmospheric chemistry, as a function of {\teff} and log(g), found in this work indicates the need to explore these effects with even finer grid spacing than this work in parameters like {\teff} (25 K in this work), log(g) (0.25 dex in this work), and {\kzz} (factor of 100 in this work). This remains the focus of a future model grid release. The model presented in this work will be updated to include self-consistent clouds in the near future. Additional future work will couple the code to the chemical kinetics model \texttt{VULCAN} \citep{tsai17,tsai21} which would enable a more rigorous self-consistent treatment of disequilibrium chemistry in our models. This will also allow us to include effects like the feedback of the disequilibrium abundances of gases like {\meth} on the entire chemical network of the atmosphere within our models, self-consistently. This will also allow us to explore the empirical trends we find in this work, like probable slower convective mixing than typically calculated assuming free convection, more robustly. Moreover, such a coupling opens up the possibility to explore both photochemistry and vertical mixing together self-consistently, which is more relevant for irradiated exoplanets. This updated model will then be used to explore effects of vertical mixing and photochemistry for warm transiting giant planets and find additional observables to constrain {\kzz} in such planets, an important comparison sample to the non-irradiated objects explored here.


\section{Acknowledgments}
SM thanks the UC Regents Fellowship award for supporting him for this work. JJF acknowledges the support of NASA XRP grant 80NSSC19K0446. We acknowledge use of the lux supercomputer at UC Santa Cruz, funded by NSF MRI grant AST 1828315. SM also thanks Caroline Morley for very useful inputs for preparing the color-magnitude diagrams presented here. We thank the anonymous referee for very insightful comments which helped in the betterment of this manuscript.
 
{\it Software:} \texttt{PICASO 3.0} \citep{Mukherjee22}, \texttt{PICASO} \citep{batalha19}, pandas \citep{mckinney2010data}, NumPy \citep{walt2011numpy}, IPython \citep{perez2007ipython}, Jupyter \citep{kluyver2016jupyter}, matplotlib \citep{Hunter:2007}, the model grid will be formally released via Zenodo.

\bibliography{arxiv}{}
\bibliographystyle{apj}



\end{document}